\documentclass[lettersize,journal]{IEEEtran}
\usepackage{amsmath,amsfonts}
\usepackage{algorithmicx}
\usepackage{algpseudocode}
\usepackage{array}
\usepackage{algpseudocode}  

\makeatletter
\newif\if@restonecol
\makeatother

\usepackage[ruled,linesnumbered]{algorithm2e}

\usepackage[caption=false,font=footnotesize,labelfont=rm,textfont=rm]{subfig}
\usepackage{textcomp}
\usepackage{stfloats}
\usepackage{url}
\usepackage{verbatim}
\usepackage{graphicx}
\usepackage{cite}
\usepackage{hyperref} 
\usepackage{url}
\usepackage{color}

\begin{document}
\bstctlcite{IEEEexample:BSTcontrol} 

\title{A GAN-based Semantic Communication \\ for Text without CSI}

\author{Jin Mao,~\IEEEmembership{Student Member,~IEEE,}
		Ke Xiong,~\IEEEmembership{Member,~IEEE,}
		Ming Liu,~\IEEEmembership{Member,~IEEE,}\\	
		Zhijin Qin,~\IEEEmembership{Senior Member,~IEEE,}
		Wei Chen,~\IEEEmembership{Senior Member,~IEEE,}	\\
		Pingyi Fan,~\IEEEmembership{Senior Member,~IEEE,} 
		and Khaled Ben Letaief,~\IEEEmembership{Fellow,~IEEE}
\thanks{This work has been submitted to the IEEE for possible publication.  Copyright may be transferred without notice, after which this version may no longer be accessible.}
\thanks{Jin Mao, Ke Xiong and, Ming Liu are with the Engineering Research Center of Network Management Technology for High Speed Railway of Ministry of Education, School of Computer and Information Technology, Beijing Jiaotong University, Beijing 100044, China, also with the Collaborative Innovation Center of Railway Traffic Safety, Beijing Jiaotong University, Beijing 100044, China, and also with the National Engineering Research Center of	Advanced Network Technologies, Beijing Jiaotong University, Beijing 100044, China. E-mail: (jinmao@bjtu.edu.cn; kxiong@bjtu.edu.cn; mingliu@bjtu.edu.cn).}

\thanks{Zhijin Qin is with the Department of Electronic Engineering, Tsinghua University, Beijing 100084, China (e-mail: qinzhijin@tsinghua.edu.cn).}

\thanks{Wei Chen and Pingyi Fan are with the Beijing National Research Center for Information Science and Technology, and the Department of Electronic Engineering, Tsinghua University, Beijing 100084, China (e-mail: wchen@tsinghua.edu.cn; fpy@tsinghua.edu.cn).}

\thanks{Khaled Ben Letaief is with the Department of Electrical and Computer Engineering, The Hong Kong University of Science and Technology (HKUST), Hong Kong, China (e-mail: eekhaled@ece.ust.hk).}
}

\markboth{Journal of \LaTeX\ Class Files,~Vol.~14, No.~8, August~2022}%
{Shell \MakeLowercase{\textit{et al.}}:A GAN-based Semantic Communication for Text without CSI}

\maketitle
\begin{abstract}
	
	Recently, semantic communication (SC) has been regarded as one of the most potential paradigms of 6G. Current SC frameworks require the physical layer channel state information (CSI) in order to handle the severe signal distortion induced by channel fading. Since practical CSI cannot be obtained accurately and the overhead of channel estimation cannot be neglected, we therefore propose a generative adversarial network (GAN) based SC framework (Ti-GSC) that doesn't require CSI. In Ti-GSC, two main modules, i.e., an autoencoder-based encoder-decoder module (AEDM) and a GAN-based non-CSI signal distortion suppression (SDS) module (GSDSM) are included where AEDM first encodes the data in the semantic dimension at the source before transmission, and then GSDSM suppresses the distortion of the received signals in both syntactic and semantic dimensions at the destination. At last, AEDM decodes the distortion-suppressed signal at the destination. As SDS only relies on learning the syntactic distribution and the semantics of the transmitted data, no prior information such as CSI is needed by GSDSM. In order to measure signal distortion, a novel loss function is proposed where two terms, i.e., a syntactic distortion loss term and a semantic distortion loss term are newly added, and a differentiable semantic measurement method is designed based on the intermediate layers of the AEDM decoder. To achieve better training results of Ti-GSC   , two training schemes, i.e., the joint optimization based training (JOT) and the alternating optimization based training (AOT) are designed for the proposed Ti-GSC. Experimental results show that JOT is more efficient for Ti-GSC, and Ti-GSC outperforms conventional communication frameworks in terms of bilingual evaluation understudy (BLEU) score in both Rician and Rayleigh fading channels. Moreover, without CSI, BLEU score achieved by Ti-GSC is about 40$\%$ and 62$\%$ higher than that achieved by existing SC frameworks in Rician and Rayleigh fading, respectively. Besides, each term of the presented loss function has great impact on the BLEU performance of Ti-GSC, where in Rician fading syntactic learning has the greatest impact, and in Rayleigh fading, the adversarial learning becomes important. 

\end{abstract}


\section{Introduction} 
\subsection{Background}

\IEEEPARstart{T}{he} sixth generation (6G) network \cite{luan20226g, 6704826, 9140412, 8334613, 10032267,10233705,10163978} is expected to provide human-centric communications \cite{dang_what_2020} and intelligent applications such as digital twins \cite{wang_survey_2023} and Metavers \cite{zhang_towards_2022}, which require high data rates. Due to the limitation of frequency spectrum, time and energy, etc., the conflict between the ever-growing user demands and the communication resources has been exacerbated. To release such a conflict, semantic communication (SC) \cite{weaver_weaver_nodate} has recently re-aroused the increasing attention of researchers.

SC incorporates semantic dimensions based on the meaning conveyed by messages \cite{EXKSC, zhou_adaptive_2022, lu_rethinking_2023, wang_knowledge_2023}. In theory, SC achieves semantic extraction via training artificial neural networks (ANN), where differentiable channels are assumed to enable back-propagation of gradients (BPG) for effective training \cite{farsad_deep_2018}. By extracting semantic information (including contextual and user relations), SC enhances the error correction capability of communication systems and also reduces redundancy of unimportant information. Unlike conventional bit-wise communications, SC encodes data by using the interrelated higher dimensional units. Therefore, it is capable of notably improving the efficiency and reliability of communications, and has been widely regarded as a potential paradigm in 6G \cite{zhang_toward_2022, yang_semantic_2022, gunduz_beyond_2022}.

\subsection{Related Work}

Up to now, SC has attracted a lot of interests. As described in \cite{weaver_weaver_nodate}, there are two levels in SC, i.e., the semantic level and the task-oriented level. At the semantic level, its main purpose is to semantically encode the data with desired meaning and decode the received signals precisely. Thereby, recovery accuracy becomes a key index at the semantic level. Comparably, the main purpose at the task-oriented level mainly focuses on influencing the actions of the receiver in the intended way. Thus, the accuracy of task execution is a measure of concern at the task-oriented level. Therefore, some works have focused on improving the transmission accuracy of SC at the semantic level see e.g. \cite{wang_performance_2022} and \cite{ dai_nonlinear_2022}, and others concentrated on enhancing the effectiveness of SC at the task-oriented level, see e.g. \cite{shao_learning_2022, shao_task-oriented_2022, xie_robust_2022, tang_contrastive_2023}.

\textit{Although, many notable results have been achieved by exiting works, they only discussed SC in additive white Gaussian noise (AWGN) channels.} As is known, the stochasticity of channel fading cannot be neglected in practice at the physical layer, so some recent works have begun to study SC over fading channels. It is noticed that, compared with AWGN channel, the ANN training for SC in fading channels is more challenging \cite{xie_lite_2021}. First, the channel fading stochasticity induces the back-propagation to local optimum during the training phase. Second, the received distorted signal on the forward-propagation may prohibit the system from recovering the semantic information accurately during both training and testing phases. Thus, how to reduce the impact of channel fading on SC requires much effort \cite{xie_deep_2021, zhou_semantic_2022, xie_task-oriented_2022, xie_lite_2021, yang_ofdm-guided_2022}. 

With perfect channel state information (CSI) assumption, some  transformer-based SC frameworks in fading channels have been presented thus far. In \cite{xie_deep_2021}, it proposed an SC framework based on transformer to extract and recover text data in fading channels. In \cite{zhou_semantic_2022}, it adopted adaptive universal transformer to flexibly handle semantic differences and adapt to different channel conditions. As perfect CSI is hard to be obtained in practice, in \cite{xie_task-oriented_2022}, it proposed single-modal and multimodal multi-user SC frameworks with estimated imperfect CSI, where it showed that imperfect CSI greatly degrades the SC's performance. Although in \cite{xie_lite_2021} and \cite{yang_ofdm-guided_2022}, they proposed a supervised-learning based method to achieve good channel estimation for SC, the pilot overhead of channel estimation still exists, and it is difficult to obtain accurate labels for supervised-learning in dynamic wireless environments.

\subsection{Motivations and Contributions}
%
%
Motivated by this, this study aims to design a novel SC framework that does not require physical layer CSI. It is noticed that, general SC framework has an encoder and a decoder. But such an encoder-decoder module is different from some known applications. Particularly, for image-based applications, the encoder-decoder module aims at extracting local features and spatial hierarchical relations, and for audio-based applications, it should be devised to extract relevant information from the time or frequency domains, for text-based applications, the encoder-decoder module needs to capture semantic relations and temporal relationships of phrase information. As text-based applications are one of the most popular ones in our daily life, in this study, we focus on design an efficient SC for text data without CSI. 
	
We note that the generative adversarial network (GAN) is capable of offering notable advantages when applied to SC. In \cite{han_semantic-preserved_2023, huang_toward_2023, jiang_wireless_2023, yue_learned_2023, zhang_semantic_2023}, they employed GAN for high fidelity reconstruction of speech, images, conference video, talking-head video and 3D human mesh, respectively. In \cite{chen_trust-worthy_2023} and \cite{zhang_deep_2022}, they applied GAN to SC for domain adaptation (DA). The aforementioned works show that GAN is competent of learning the data distribution of input information, enabling the mapping of input information and output information. In view of this, we propose a new SC framework by integrating GAN with SC, as the inherent generative capacity of GAN can be used to generate original semantic signal from distorted semantic signal, and thus CSI is not required. The main contributions of this paper are summarized as follows:

\begin{figure}[!t]
\centering
\includegraphics[width=3.4in]{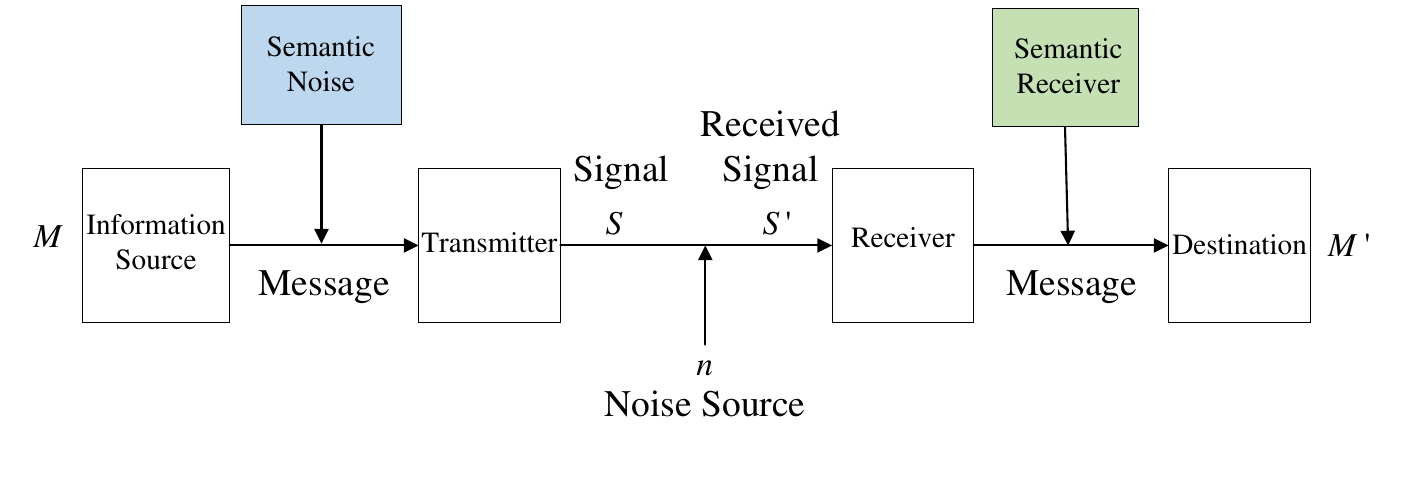}
\caption{Semantic communication Framework proposed by Weaver.}
\label{fig_sc_system}
\end{figure}

\begin{itemize}

	\item {A novel GAN-based SC framework for text input (Ti-GSC), which doesn't require CSI, is presented, where we integrate GAN at the receiver to assist the semantic decoder to better decoding the received data. The proposed Ti-GSC consists of an autoencoder-based encoder-decoder module (AEDM) to encode as well as decode data in semantic dimension and a GAN-based non-CSI signal distortion suppression (SDS) module (GSDSM). In GSDSM, two more neural networks are cooperated with AEDM. One is used to suppress the received distorted signal in both syntactic and semantic dimensions, and the other is to train the first one. Due to the fact that GAN can map the received signal to the distribution of the transmitted signal, it can generate signals syntactically and semantically similar to the transmitted signal. Therefore, CSI is not required.}

	\item {To efficient measure the distortion, two terms are newly added to the total loss function, and a differentiable semantic measurement method is devised. Two terms are used to measure the signal syntactic distortion and semantic distortion, respectively. The differentiable semantic measurement method is used to perceive the semantic difference.}

	\item To better understand the proposed framework, we discuss two distinct training schemes. The first scheme is optimizing AEDM and GSDSM jointly, i.e., joint optimizing based training (JOT), and the second scheme is optimizing them alternatively, i.e., alternative optimizing based training (AOT). The experimental results demonstrate that our proposed framework outperforms conventional frameworks without CSI in terms of bilingual evaluation understudy (BLEU) scores in both Rician fading and Rayleigh fading channels. Notably, the BLEU score is enhanced by approximately 40$\%$ and 62$\%$ when GSDSM is incorporated into the presented Ti-GSC framework, compared to that without utilizing GSDSM, in Rician and Rayleigh fading channels, respectively. Furthermore, the ablation experiment confirmed the importance of each term in the GSDSM.
		 		\begin{figure}[ht]   
		 	\centering

		 	\captionsetup[subfloat]{singlelinecheck=off}
		 	\begin{minipage}{0.49\textwidth} 

		 		\subfloat[The traditional SC framework. The green part described in the dashed line is the AEDM.]{\includegraphics[width=1\textwidth]{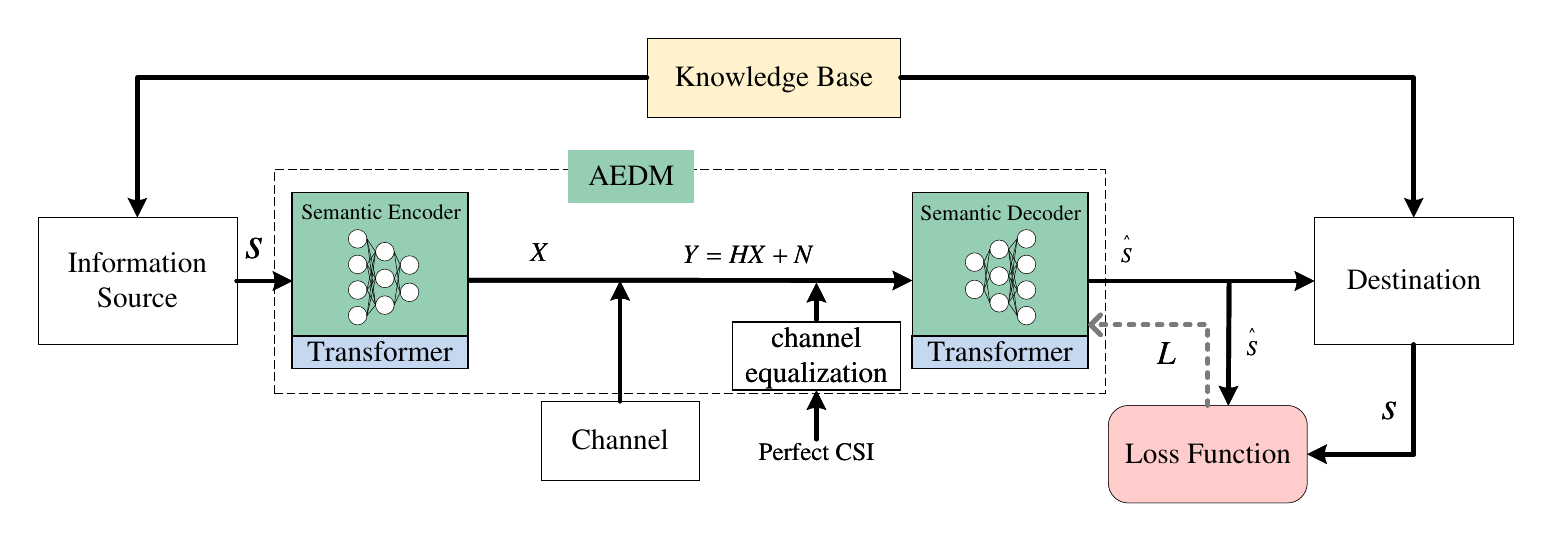}\label{fig_sysmod_1}}			
		 	\end{minipage}
		 	\vfill
		 	\begin{minipage}{0.49\textwidth} 						
		 		\subfloat[The proposed SC framework. The green and red parts described in the dashed line are the proposed Ti-GSC framework.]{\includegraphics[width=1\textwidth]{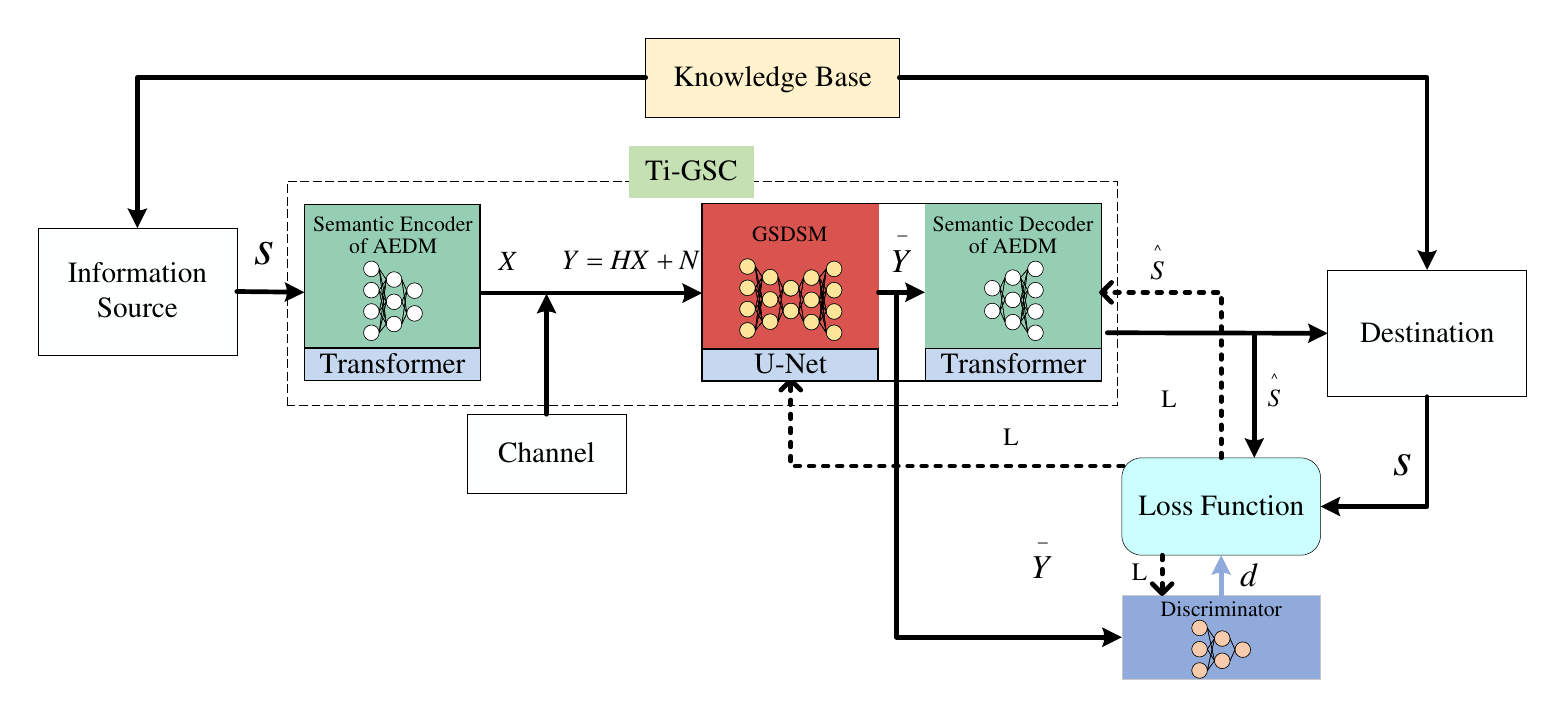}\label{fig_sysmod_2}}
		 		\caption{The SC frameworks of traditional one and proposed one.}	
		 	\end{minipage}
		 \end{figure}
	\item To evaluate the function of AEDM by means of a clear and understandable visual representation, t-SNE algorithm is adopted to visualize the spatial distribution of encoded signals to explore the mechanism of AEDM. The visualization diagram shows that ADEM encodes signal in two layers, where distinct large clusters correspond to different words, while within a given large cluster, smaller clusters emerge within the larger cluster due to inherent semantic variations. The robustness of our proposed framework is also demonstrated by comparing the visualization representation of received signals and distortion suppression signals under different signal-to-noise ratios (SNR). 
	
	\item To the best of our knowledge, this is the first work leveraging GAN into text SC for suppressing signal distortion without additional prior information such as CSI, and realizing a differentiable self-measurement method for semantic distortion measurement without the aid of other pre-trained models (e.g. BERT and VGG \footnote{Bidirectional Encoder Representations from Transformers (BERT) model is designed for Text \cite{devlin_bert_2019} and Visual Geometry Group network (VGG) is designed for images \cite{Simonyan:2014cmh}.}) to assist in decoder learning. 
\end{itemize}

\subsection{Organizations}
	The rest of this paper is organized as follows. Section \uppercase\expandafter{\romannumeral2} describes the traditional SC framework and the proposed Ti-GSC framework including AEDM and GSDSM. Section \uppercase\expandafter{\romannumeral3} introduces the two training schemes of the the proposed Ti-GSC framework. Section \uppercase\expandafter{\romannumeral4} provides a detailed explanation of the experiment settings and the results. We conclude it in Section \uppercase\expandafter{\romannumeral5}.

\subsection{Notations}
	Boldface lower-case letters and upper-case letters denote vectors and matrices, respectively. $\left| \cdot \right|$ denotes the modulus. $w_{i}\in\boldsymbol{s}$ means the $i$-th element of vector $\boldsymbol{s}$. $\mathbb{E}$ denotes the expectation operator.
	$ \left\| \cdot \right\| $ denotes the Euclidean norm. $\nabla$ denotes the gradient operator. $\boldsymbol{x}\sim\mathcal{CN}(0,\sigma^{2})$ means the variable $\boldsymbol{x}$ follows a circularly symmetric complex Gaussian distribution with mean 0 and covariance $\sigma^{2}$.
\section{Proposed GAN-Based Non-CSI Semantic Communication Framework}

\subsection{Review of General Semantic Communication Framework for Text Input}
	The traditional framework of SC for text input is shown in Fig. 2(a), where both the information source and destination share a common knowledge base. Denote a sentence $\boldsymbol{s}$ with word number $\left| \boldsymbol{s} \right| = N$ as the input data. A complete SC process starts with the information source sending information $\boldsymbol{s}$ and ends with the destination obtaining the decoded information $\hat{\boldsymbol{s}}$. As shown in Fig. 2(a), $\boldsymbol{s}$ is the input of the semantic encoder, and the encoded signal $\boldsymbol{X}$ is the output of semantic encoder. After channel transmission, the received signal $\boldsymbol{Y}$ is first equalized by inputting CSI and then decoded by semantic decoder to recover $\hat{\boldsymbol{s}}$. For such a SC framework, it first requires to train the semantic encoder and the semantic decoder, and then one can obtain $\hat{\boldsymbol{s}}$ by inputting $\boldsymbol{s}$. In the training phase, both $\boldsymbol{s}$ that stored in the knowledge base and $\hat{\boldsymbol{s}}$ are inputted for a designed loss function to gradually update the parameters of semantic encoder and the semantic decoder. As marked in Fig. 2(a), the semantic encoder and the semantic decoder are also referred to as AEDM.
	
	AEDM can be a typical ANN \cite{xie_deep_2021}, except for removing the assumption of obtaining perfect CSI for channel equalization and the mutual information model. For clarity, the AEDM is depicted in green in Fig. 2(b), where the transformer \cite{transformer_2017} is employed as the core structure for AEDM, which extracts the semantic information by using self-attention mechanism. Semantic encoder defined as $ E_{\rm nn}^{\rm encd}(\cdot )$ with parameters $\boldsymbol{W_{e}}$ encodes data in the semantic dimension. Sentence $\boldsymbol{s}$ is inputted into $E_{\rm nn}^{\rm encd}(\cdot )$, then the transmitted signal $\boldsymbol{X}$ is delivered in the fading channel with channel gain denoted by $\boldsymbol{H}$, where $\boldsymbol{X} = E_{\rm nn}^{\rm encd}(\boldsymbol{s}|\boldsymbol{W_{e}}) $. Thus the received signal $\boldsymbol{Y}$ is given by
	 
	\begin{equation}
		\label{eq:rciv signal}
		\boldsymbol{Y} = \boldsymbol{H}\boldsymbol{X}+\boldsymbol{N},
	\end{equation}
	where $\boldsymbol{N}\sim\mathcal{CN}(0,\sigma^{2}) $ is the AWGN. Once received $\boldsymbol{Y}$, semantic decoder defined as $D_{\rm nn}^{\rm decd}(\cdot )$ with parameters $\boldsymbol{W_{d}}$ decodes $\hat{\boldsymbol{s}}$ from $\boldsymbol{Y}$, where the sentence $\hat{\boldsymbol{s}} = D_{\rm nn}^{\rm decd}(\boldsymbol{Y}|\boldsymbol{W_{d}})$.

	The loss function of AEDM is considered as Cross Entropy (CE) loss, which is given by
	\begin{equation}
		\label{eq:cross entroyl}
		L_{\rm CE}=-\sum_{i=1}^{\rm N}p(w_{i})\log(q(w_{i})),
	\end{equation}
	where $p(w_{i})$ and $q(w_{i})$ represent the the real probability and the predicted probability of the $i$-th word $w_{i}$ that appears in sentence $\boldsymbol{s}$ and $\hat{\boldsymbol{s}}$, respectively.

\subsection{Proposed GAN-based Non-CSI Semantic Communication Framework}
	The semantic encoder and semantic decoder of Ti-GSC are the same as AEDM in traditional SC framework illustrated in Fig. 2(a). The main difference between our presented Ti-GSC and the traditional SC framework is that a module called GSDSM is added at the receiver to incorporate with AEDM. As shown in Fig. 2(b), $\boldsymbol{Y}$ is first processed by the GSDSM to obtain $\overline{\boldsymbol{Y}}$, which is then decoded by $D_{\rm nn}^{\rm decd}(\cdot )$ to recover $\hat{\boldsymbol{s}}$. 
		
	Specifically, GSDSM is adopted to learn the semantic mapping between the received and the transmitted signals, enabling distortion suppressing of the received signal for better decoding of AEDM at the receiver. Without CSI, the correlation of semantics of the encoded signal is learned and used by GSDSM to generate the distortion suppressed signal that are semantically similar to the encoded signal from the received signal against semantic noise. Meanwhile, the channel characteristics can be implicitly learned from the input-output mapping of GSDSM to against the channel fading effect.

	GSDSM can be realized by a GAN \cite{goodfellow_generative_2014} involving a discriminator to implement adversarial learning. The reasons can be expressed as follows: First, semantic information embedded in the encoded signal can be effectively leveraged by GAN for signal distortion suppression (SDS). Second, both GAN and SC employ  gradient back-propagation method for updates, making joint optimization a natural advantage. Third, as described in \cite{shannon_mathematical_nodate}, depicted in Fig. \ref{fig_sc_system}, apart from channel effects, the transformation of source information into the transmitted signal introduces semantic noise that adversely impacts the decoding of the destination. It requires  semantic decoder at the receiver to adjust the semantic characteristics of the received signal to match the semantic capacities of the decoder. By leveraging GAN, it is possible to adjust the received signal to align with the semantic capabilities of the decoder. 
	
	In order to efficiently train the GSDSM, except for the CE loss function of AEDM, corresponding adversarial loss functions of GSDSM and discriminator are required, i.e., the adversarial loss of GSDSM and the adversarial loss of discriminator. Moreover, due to the presence of channel fading and semantic noise, the performance of Ti-GSC maybe affected. In order to reduce the influence of channel effects and semantic noise by adjusting the signals generated by GSDSM to preserve syntactic and semantic information, two new terms were added to the total loss function, namely syntactic loss function and semantic loss term. We give the details of GSDSM (including discriminator) and designed loss functions bellow.

	Before introducing GSDSM, the general framework of GAN is described first. The parameters of the generator $G(\cdot)$ and discriminator $D(\cdot)$ are defined by $\pmb{\Phi}_{g}$ and $\pmb{\Phi}_{d}$, respectively. $G(\boldsymbol{z}|\pmb{\Phi}_{g})$ generates data corresponding to the distribution $p_{g}$ of training data $\boldsymbol{x}$, where the noise $\boldsymbol{z}\sim p_{\boldsymbol{z}}(\boldsymbol{z})$ and $p_{\boldsymbol{z}}$ is usually a Gaussian distribution. $D(\cdot|\pmb{\Phi}_{d})$ discriminates between real data and generated data. The main objective of $G(\boldsymbol{z}|\pmb{\Phi}_{g})$ is to minimize the correct discrimination possibility of $D(\boldsymbol{x}|\pmb{\Phi}_{d})$, and for $D(\boldsymbol{x}|\pmb{\Phi}_{d})$ is to maximize the possibility of identifying real and fake data.	The two-player min-max optimization problem with value function can be formulated to be

    \begin{equation}
    \begin{aligned}
    \label{eq:op_of_gan_van}
    \min \limits_{\pmb{\Phi}_{g}} \max \limits_{\pmb{\Phi}_{d}} V(D,G)= 
    \mathbb{E}_{{\boldsymbol{x} \sim p_{data}}}[\log(D(\boldsymbol{x}|\pmb{\Phi}_{d}))] \\
    + \mathbb{E}_{{\boldsymbol{z} \sim p_{\boldsymbol{z}}}}[\log(1-D(G(\boldsymbol{z}|\pmb{\Phi}_{g})|\pmb{\Phi}_{d}))].
    \end{aligned}
    \end{equation}

     \begin{figure}[!t]
     	\centering
     	\includegraphics[width=3.6in]{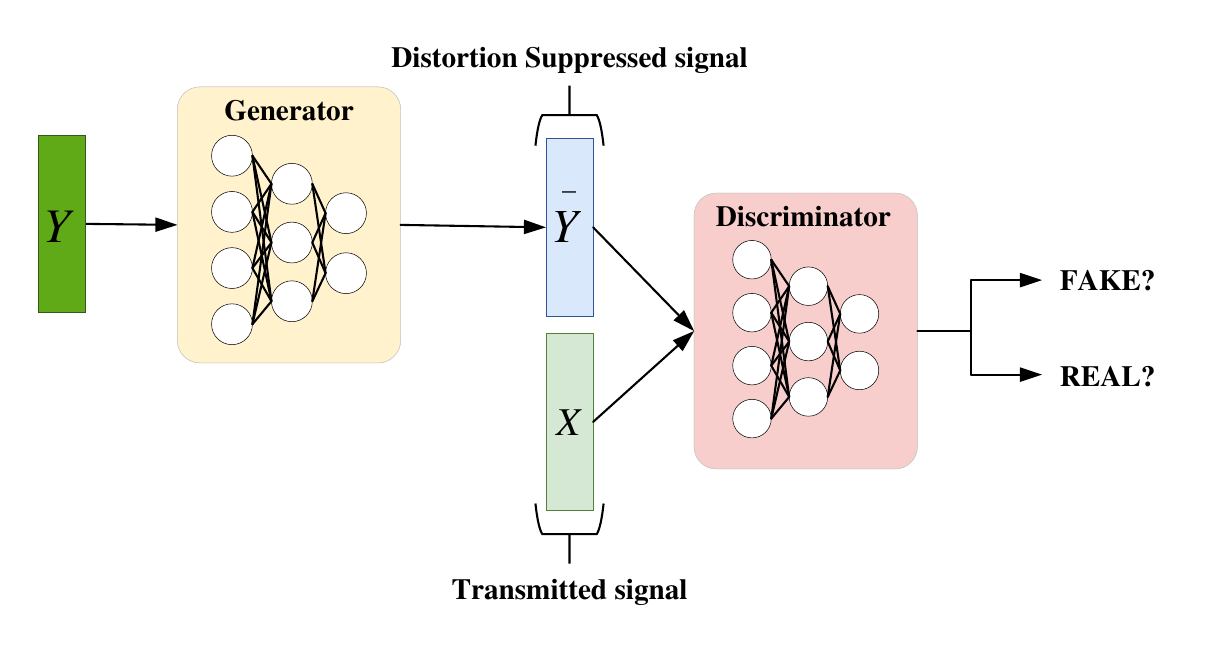}
     	\caption{The principle illustration of GSDSM.}
     	\label{fig_gan}
     \end{figure}
     Based on GAN, the proposed GSDSM including discriminator is introduced. The parameters of GSDSM $G_{\rm nn}^{\rm gnrt}(\cdot )$ which is portrayed in red in Fig. 2(b), and discriminator $D_{\rm nn}^{\rm dscm}(\cdot )$ which is portrayed in blue in Fig. 2(b) are defined by $\pmb{\Theta}_{d}$ and $\pmb{\Theta}_{g}$, respectively. One can train $G_{\rm nn}^{\rm gnrt}(\cdot )$ to map the original transmission signal from the distorted signal\footnote{Through experimental verification, it is better to train our model to predict undistorted data rather than predicting distortion values, which is consistent with the well-known generative work DALL$\cdot$E 2\cite{ramesh_hierarchical_2022}.}, which confirms $\overline{\boldsymbol{Y}} = G_{\rm nn}^{\rm gnrt}(\boldsymbol{Y}|\pmb{\Theta}_{g})$, where $\overline{\boldsymbol{Y}}$ denotes the distortion-suppressed signal. $D_{\rm nn}^{\rm dscm}(\cdot )$ discriminates between distortion suppressed data and transmitted data, the principle of suppressing signal distortion is shown in the Fig. \ref{fig_gan}. Therefore, the reformatted optimization problem\footnote{During the wireless transmission, AWGN is introduced to the received signal $\boldsymbol{y}$. Therefore, there is no requirement to input additional noise $\boldsymbol{z}$ as indicated in \eqref{eq:op_of_gan_van}.} is 
     \begin{equation}
    \begin{aligned}
    \label{eq:op_of_gan_c}
    \min \limits_{\pmb{\Theta}_{g}} \max \limits_{\pmb{\Theta}_{d}} V(D_{\rm nn}^{\rm dscm},G_{\rm nn}^{\rm gnrt})= 
    \mathbb{E}_{{\boldsymbol{X} \sim p_{\boldsymbol{X}}}}[\log(D_{\rm nn}^{\rm dscm}(\boldsymbol{X}|\pmb{\Theta}_{d}))] \\
    + \mathbb{E}_{{\boldsymbol{Y} \sim p_{\boldsymbol{Y}}}}[\log(1-D_{\rm nn}^{\rm dscm}(G_{\rm nn}^{\rm gnrt}(\boldsymbol{Y}|\pmb{\Theta}_{g}))|\pmb{\Theta}_{d})].
    \end{aligned}
    \end{equation}
    \IEEEpubidadjcol
    
    To train the GAN stably, WGAN-gp that employs wasserstein distance and gradient penalty (gp) is adopted. The overall optimization problem is updated by
    \begin{equation}
    \begin{aligned}
    \label{eq:op_of_gan_gp}
    \min \limits_{\pmb{\Theta}_{g}} \max \limits_{\pmb{\Theta}_{d}} V(D_{\rm nn}^{\rm dscm},G_{\rm nn}^{\rm gnrt})= 
    -\mathbb{E}_{{\boldsymbol{X} \sim p_{\boldsymbol{X}}}}[D_{\rm nn}^{\rm dscm}(\boldsymbol{X}|\pmb{\Theta}_{d})]  \\
    +\mathbb{E}_{{\boldsymbol{Y} \sim p_{\boldsymbol{Y}}}}[D_{\rm nn}^{\rm dscm}(G_{\rm nn}^{\rm gnrt}(\boldsymbol{Y}|\pmb{\Theta}_{g})|\pmb{\Theta}_{d})]\\[1mm]
     + \lambda\mathbb{E}_{{\boldsymbol{\hat{Y}_{d}} \sim
     p_{\boldsymbol{\hat{Y}_{d}}}}}[(\left\|\nabla_{\boldsymbol{\hat{Y}_{d}}}D_{\rm nn}^{\rm dscm}(\boldsymbol{X}|\pmb{\Theta}_{d})\right\|_2-1)^{2}],
    \end{aligned}
    \end{equation}
    where $\boldsymbol{\hat{Y}_{d}}$ is random interpolation sampling between the transmitted signal $\boldsymbol{X}$  and the generated signal $\overline{\boldsymbol{Y}}$. Following (5), the adversarial loss function of $G_{\rm nn}^{\rm gnrt}$  is given by
     \begin{equation}
    	\label{eq:adverasrial_loss_g}
    	L_{\rm adv\_g}=-\mathbb{E}[\left\|D_{\rm nn}^{\rm dscm}(\overline{\boldsymbol{Y}}|\pmb{\Theta}_{d}) \right\|_2^{2}].
    \end{equation}
    The adversarial loss function of $D_{\rm nn}^{\rm dscm}$ is denoted by
     \begin{equation}
     \begin{aligned}
   	\label{eq:adverasrial_loss_d}
   	L_{\rm adv\_d}=\mathbb{E}\left[
   	\left\|D_{\rm nn}^{\rm dscm}(\overline{\boldsymbol{Y}}|\pmb{\Theta}_{d})\right\|_2^{2}
   	-\left\|D_{\rm nn}^{\rm dscm}(\boldsymbol{Y}|\pmb{\Theta}_{d})\right\|_2^{2} \right.\\
   	\left.+\lambda\left\|(\left\|\nabla_{\boldsymbol{\hat{Y}_{d}}}D_{\rm nn}^{\rm dscm}(\boldsymbol{X}|\pmb{\Theta}_{d})\right\|_2-1)^{2} \right\|_2^{2}  \right].
   	\end{aligned}
    \end{equation}
    
    Recall that, there are two kinds of negative impacts existing in the SC framework. One is the channel effect (also referred to as low-dimensional distortion) which is induced by white noise and channel fading etc., and the other is semantic noise (also referred to as high-dimensional distortion) which causes different semantic similarity between the input and the output. In order to reduce the impact of channel effect and semantic noise, several distortion measurements have been presented, see e.g. \cite{liu_rate-distortion_2021} and \cite{sagduyu_is_2022}. In \cite{liu_rate-distortion_2021}, intrinsic state (IS) and extrinsic observation (EO) are used to capture distortion.  In \cite{sagduyu_is_2022}, a customized loss consisting of reconstruction loss (Mean Squared Error (MSE) loss) and classifier loss (CE loss) was proposed to train the system. But IS can be the classification label at the task-oriented level, which is not incompatible with our task at the semantic level. The customized loss is adopted to address the low-dimensional distortion, however, in \cite{sagduyu_is_2022}, high dimensional distortion is ignored in their loss function. As high dimensional distortion cannot be neglected in SC, in this study we employ a measurement method to estimate both low dimension distortion and high dimension distortion, defined by
    \begin{equation}
    \label{eq:distortion_difination}
    R(\boldsymbol{X}, \overline{\boldsymbol{Y}})= \mathbb{E}[D_{\rm  sytc}(\overline{\boldsymbol{Y}},\boldsymbol{Y})]+ \mathbb{E}[D_{\rm smtc}(\overline{\boldsymbol{Y}},\boldsymbol{Y})],
    \end{equation}
   where $D_{\rm sytc}(\overline{\boldsymbol{Y}},\boldsymbol{Y})$ and $D_{\rm smtc}(\overline{\boldsymbol{Y}},\boldsymbol{Y})$ represent the measurement of low-dimensional distortion (referred to as syntactic distortion loss function) and high-dimensional distortion (referred to as semantic distortion loss function), respectively.

   For the term of $D_{\rm sytc}(\overline{\boldsymbol{Y}},\boldsymbol{Y})$, we set it as the 2-norm to measure the reconstructive difference between transmitted signal and remodeled signal, and the syntactic distortion loss function $L_{\rm sytc}$ is expressed by
    \begin{equation}
    \label{eq:l2_loss}
    L_{\rm sytc} = \mathbb{E}[D_{\rm sytc}(\cdot )]=\mathbb{E}[\left\|\boldsymbol{X} - G_{\rm nn}^{\rm gnrt}(\boldsymbol{Y}|\pmb{\Theta}_{g}) \right\|_2^{2}].
    \end{equation}

	{For the term of $D_{\rm smtc}(\overline{\boldsymbol{Y}},\boldsymbol{Y})$, as semantic distortion cannot be measured directly and differentiable metrics are necessary for ANN updates, an effective way to map signals to differentiable semantic information representation is required. Thus in \cite{huang_deep_2021}, a pre-trained VGG network was used as a mapping function to extract image features as a feature level loss. However, using a pre-trained model VGG may add some uncontrollable factors \cite{lu_reinforcement_2022}, in order to measure the semantic difference in a differential and controllable manner, in this research, we set a portion of the semantic decoder as the mapping function of the semantic distortion loss function $D_{\rm smtc}$ to perceive the semantic differences between the transmitted signal and the distortion-suppressed signal in the latent representation. The semantic distortion loss function $L_{\rm smtc}$ is defined by}    
    \begin{equation}
    \begin{aligned}
    \label{eq:smtc_loss}
    L_{\rm smtc}&=\mathbb{E}[D_{\rm smtc}(\cdot )] \\
    &=\mathbb{E}[\left\|f(\boldsymbol{X}) - f(G_{\rm nn}^{\rm gnrt}(\boldsymbol{Y}|\pmb{\Theta}_{g})) \right\|_2^{2}],
	\end{aligned}
    \end{equation}
    where $f(\cdot )$ represents intermediate layers of the semantic decoder. Taking \eqref{eq:l2_loss} and \eqref{eq:smtc_loss} to \eqref{eq:distortion_difination}, we have the distortion loss function $L_{\rm dstr}$, defined by
    \begin{equation}
    \label{eq:distortion_loss}
    L_{\rm dstr}=\lambda * L_{\rm sytc}+\gamma * L_{\rm smtc},
    \end{equation}
    where $ \lambda $ and $ \gamma $ are the weights of syntactic loss and semantic loss, respectively. By adding the distortion loss to adversarial learning, we restore both the syntactic information and semantic information.

\section{Training Schemes of GAN-based Non-CSI Semantic Communication Framework}
	For Ti-GSC described in Section \uppercase\expandafter{\romannumeral2}.B, the training process is extremely important, as two ANN modules are adopted to realize the AEDM and GSDSM. As is known, there are two popular training schemes i.e., JOT and AOT, to train the neural networks. In order to find the more proper training method for Ti-GSC, both AOT and JOT are designed for training Ti-GSC. In JOT, as shown in Fig. 4(a), GSDSM and AEDM are considered as an entire framework. By performing multiple rounds of updates, GSDSM and AEDM can be trained jointly with JOT. In AOT, as depicted in Fig. 4(b), three steps are included. First, GSDSM including discriminator is updated. Second, the AEDM with GSDSM is updated. At last, the whole framework is updated by using JOT. By taking turns executing the three steps, Ti-GSC can be well trained with AOT.
	
	\begin{figure}[htbp]   
		\centering
		\captionsetup[subfloat]{singlelinecheck=off}
			\begin{minipage}{0.49\textwidth}
			\vspace{-0.5cm} 
			\centering	
			\subfloat[The training flow of proposed framework in JOT.]{\includegraphics[width=1\textwidth]{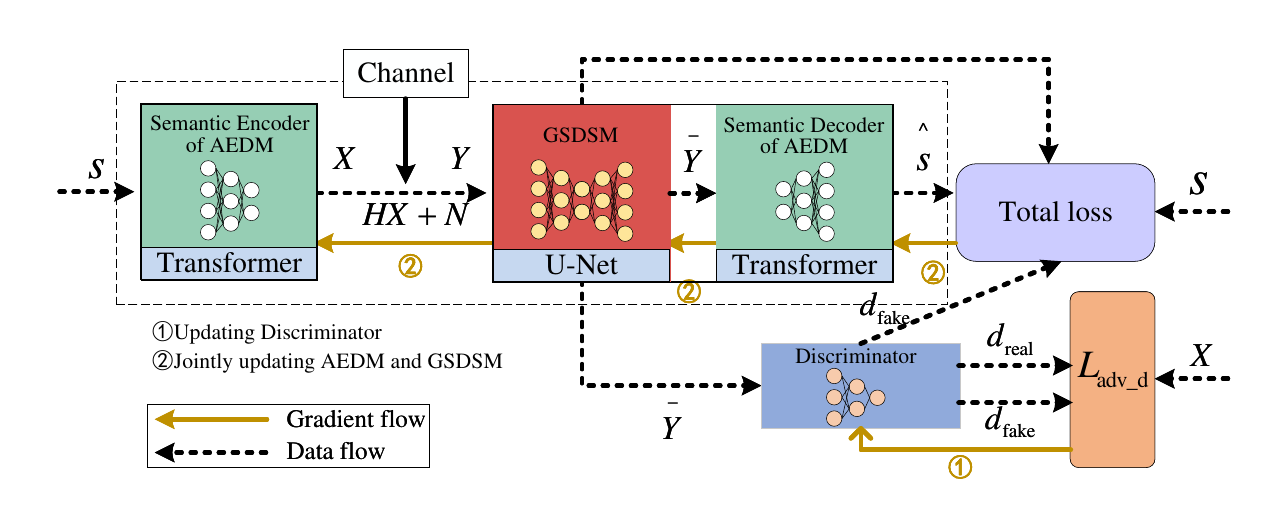}
				\label{fig_ts2}}
			
		\end{minipage}
		\begin{minipage}{0.49\textwidth} 
			\vspace{-0.1cm}
			\centering	
			\subfloat[The training flow of proposed framework in AOT.]{\includegraphics[width=1\textwidth]{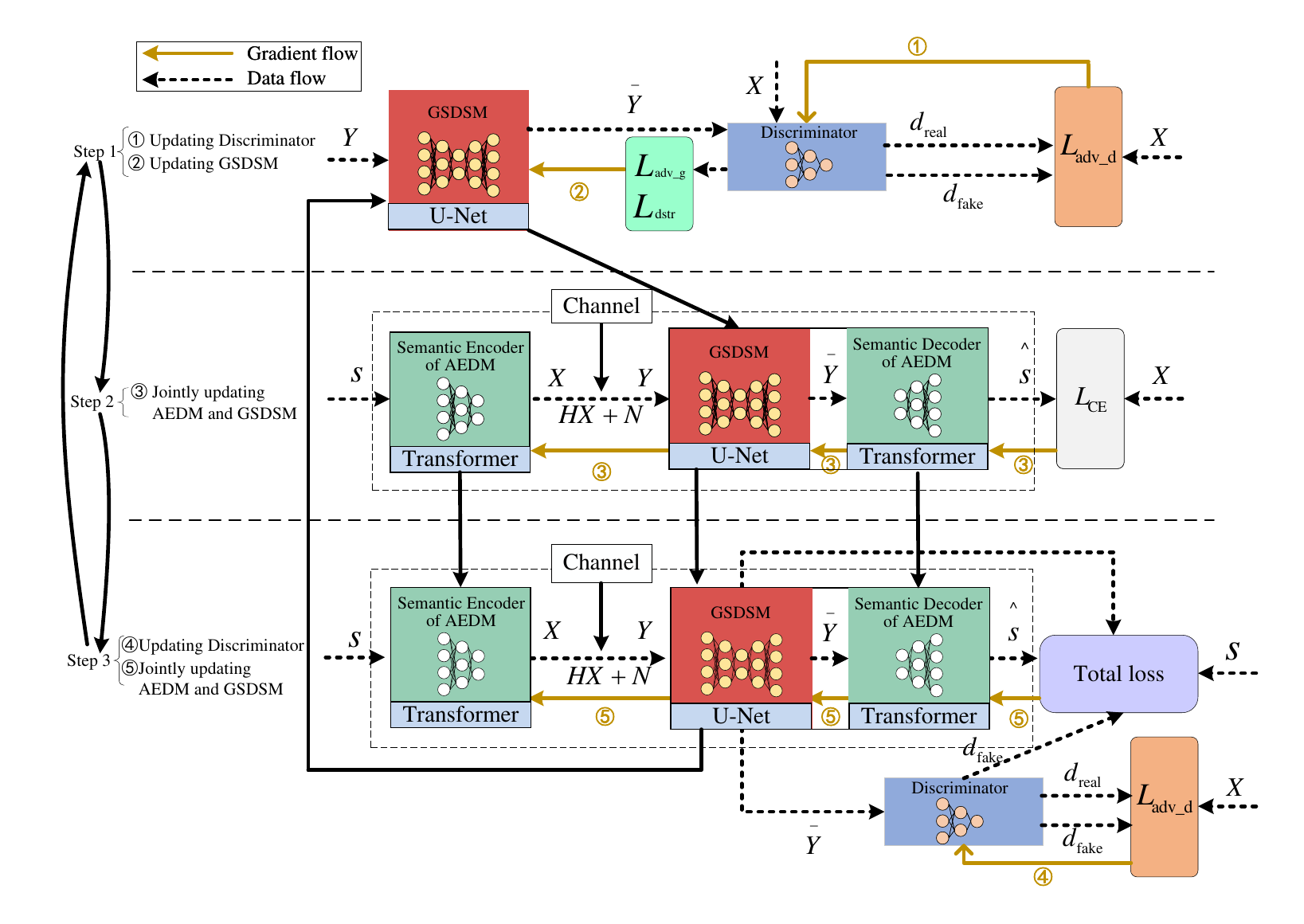}
				\label{fig_ts1}}
			\caption{The training flow of two training schemes.}
		\end{minipage}
		
	\end{figure}

   \subsection{JOT}
	    
	In JOT, AEDM and GSDSM are regarded as an entire neural network (technically, in the identical computational graph). Therefore, AEDM and GSDSM are updated by a total loss function $L_{\rm total}$, denoted by
	\begin{equation}
		\label{eq:total_loss}
		L_{\rm total} = w_{1} *L_{\rm CE} +w_{2} * L_{\rm adv\_g} +  \lambda * L_{\rm sytc}+\gamma * L_{\rm smtc},
	\end{equation}	
	where $w_{1}$ and $w_{2}$ are the weights of $L_{\rm CE}$ and $L_{\rm adv\_g}$, respectively.
	
	As shown in Fig. 4(a), the discriminator is updated by computing $L_{\rm adv\_d}$. Then, AEDM and GSDSM are updated simultaneously by computing the total loss function $L_{\rm total}$, see in the pseudo code \hyperlink{Algo1-ref}{Algorithm 1} for clarity. Specifically, for each epoch, and for each batch, inputting the sentence $\boldsymbol{s}_j \in \left\{\boldsymbol{s}_1, \boldsymbol{s}_2, ... , \boldsymbol{s}_B \right\}$ generated by the knowledge base $\mathcal{K}$ into the semantic encoder, where $B$ is the batch size, and transmitting the encoded signal $X$ through the channel. GSDSM receives the distorted signal $\boldsymbol{Y}$ and generate the distortion-suppressed data $\overline{\boldsymbol{Y}}$. Then, computing the random interpolation $inter$ between real and fake samples. Inputting the interpolation $inter$, transmitted signal $\boldsymbol{X}$ and distortion suppression signal $\overline{\boldsymbol{Y}}$ into $D_{\rm nn}^{\rm dscm}$ to obtain the discriminator values $d_{\rm inter}$, $d_{\rm real} $ and $d_{\rm fake}$, respectively. Next, the semantic decoder performs decoding as input the distortion suppression signal $\overline{\boldsymbol{Y}}$. Eventually, updating $D_{\rm nn}^{\rm dscm}$ by computing \eqref{eq:adverasrial_loss_d} and updating AEDM with GSDSM by computing the total loss $L_{\rm total}$.

\hypertarget{Algo1-ref}{}
\begin{algorithm}
	\caption{Training the whole framework: JOT.}
	\KwIn{The knowledge base $\mathcal{K}$.}
	\KwOut{The trained network parameters of the AEDM $W_{e}$ and $W_{d}$. The trained network parameters of the GSDSM and the discriminator $\boldsymbol{\Theta}_g$ and $\boldsymbol{\Theta}_d$, respectively.}
	\While(\hfill\tcp*[h]{Epoch size $e$}){Epoch $e$ is not zero}{
		\For(\hfill\tcp*[h]{Batch size $B$ }){$j=1,...,B$}
		{$\boldsymbol{X} = E_{\rm nn}^{\rm encd}(\boldsymbol{s_j}|W_{e})$\;
			Sending $\boldsymbol{x}$ to the AWGN, Rician fading \leavevmode\newline
			or Rayleigh fading channels\;
			GSDSM receives the distorted signal $Y$\;
			\tcp{suppressing the distorted signal}
			$\overline{Y} \gets G_{\rm nn}^{\rm gnrt}(Y|\boldsymbol{\Theta}_{g})$\;
			$d_{inter} \gets D_{nn}^{dscm}(inter|\boldsymbol{\Theta}_d)$\; 
			$d_{real} \gets D_{nn}^{dscm}(\boldsymbol{X}|\boldsymbol{\Theta}_d)$\;
			$d_{fake} \gets D_{nn}^{dscm}(\overline{\boldsymbol{Y}}|\boldsymbol{\Theta}_d)$\;
			Updating Discriminator by computing \eqref{eq:adverasrial_loss_d}\;
			Computing $L_{adv\_g}$ by \eqref{eq:adverasrial_loss_g}\;
			$\hat{\boldsymbol{s}} = D_{\rm nn}^{\rm decd}(\overline{Y}|W_{d})$\;
			Computing $L_{\rm CE}$ by (4)\;
			Computing $L_{\rm sytc}$ and $L_{\rm smtc}$ by \eqref{eq:distortion_loss}\;
			$L_{\rm total} = w_{1} *L_{\rm CE} +w_{2} * L_{\rm adv\_g} \leavevmode\newline
			+ \lambda * L_{\rm sytc}+\gamma * L_{\rm smtc}$ \;
			Updating the overall network by $L_{\rm total}$\;
		}	
	}
\end{algorithm} 
   
   \subsection{AOT}	
	To accomplish AOT, we put AEDM and GSDSM in different computational graphs and then put them in the identical computational graph as JOT. As shown in Fig. 4 (b), AOT includes three main steps: First, the discriminator is updated by computing $L_{\rm adv_d}$, and GSDSM is updated by computing $L_{\rm adv\_g}$ and $L_{\rm dstr}$. Second, AEDM with GSDSM are updated by $L_{\rm CE}$. Third, updating the whole framework by calling Algorithm 1 (JOT).
	
	\textit{a. Training GSDSM and Discriminator}
	
	As shown in the pseudo code of \hyperlink{Algo2-ref}{Algorithm 2}, GSDSM generates the distortion-suppressed signal $\overline{\boldsymbol{Y}}$ after inputting the received distorted signal $\boldsymbol{Y}$. Then, updating $D_{\rm nn}^{\rm dscm}$ by computing \eqref{eq:adverasrial_loss_d}. Lastly, computing the adversarial loss of $G_{\rm nn}^{\rm gnrt}$ ($L_{\rm adv\_g}$), and the distortion loss $L_{\rm dstr}$ by computing \eqref{eq:adverasrial_loss_g} and \eqref{eq:distortion_loss}, respectively. Update $G_{\rm nn}^{\rm gnrt}$ by computing $L_G$, where $w_{3}$ is the weight of the adversarial loss of $G_{\rm nn}^{\rm gnrt}$. 
	
	\hypertarget{Algo2-ref}{}
	\begin{algorithm}
		\caption{Training GSDSM and discriminator.}
		\KwIn{The received signal $\boldsymbol{Y}$.}
		\KwOut{The trained network parameters of the GSDSM and the discriminator $\boldsymbol{\Theta}_g$ and $\boldsymbol{\Theta}_d$, respectively.}
		Sampling $\alpha \sim$ Uniform[0,1]\;
		$\overline{\boldsymbol{Y}} \gets G_{\rm nn}^{\rm gnrt}(\boldsymbol{Y}|\boldsymbol{\Theta}_{g})$\;
		$inter \gets \alpha \boldsymbol{X} + (1-\alpha)\overline{\boldsymbol{Y}}$\;
		\tcp{for computing gp}
		$d_{inter} \gets D_{\rm nn}^{\rm dscm}(inter|\boldsymbol{\Theta}_d)$\; 
		$d_{real} \gets D_{\rm nn}^{\rm dscm}(\boldsymbol{X}|\boldsymbol{\Theta}_d)$\;
		$d_{fake} \gets D_{\rm nn}^{\rm dscm}(\overline{\boldsymbol{Y}}|\boldsymbol{\Theta}_d)$\;
		Updating Discriminator by computing \eqref{eq:adverasrial_loss_d}\;
		Computing the adversarial loss of $G_{\rm nn}^{\rm gnrt}$ ($L_{\rm adv\_g}$) by \eqref{eq:adverasrial_loss_g}\;
		Computing the semantic distortion loss $L_{\rm dstr}$ by \eqref{eq:distortion_loss}\;
		Computing $L_{\rm G}$ = $w_{3} * L_{\rm adv\_g} +L_{\rm dstr}$\;
		Updating the $G_{\rm nn}^{\rm gnrt}$ by $L_{\rm G}$\;	
	\end{algorithm}

	\textit{b. Training AEDM with GSDSM}
	
	GSDSM receives the distorted signal $\boldsymbol{Y}$ and generate the distortion-suppressed data $\overline{\boldsymbol{Y}}$ as the input into the semantic decoder, and the semantic decoder performs decoding. Updating AEDM with GSDSM by computing $L_{\rm CE}$. The pseudo code is shown in \hyperlink{Algo3-ref}{Algorithm 3}.

	\hypertarget{Algo3-ref}{}
	\begin{algorithm}
		\caption{Training AEDM with GSDSM}
		\KwIn{The received signal $\boldsymbol{Y}$.}
		\KwOut{The network parameters of AEDM $W_{e}$ and $W_{d}$. The network parameters of GSDSM $\boldsymbol{\Theta}_g$.}
		$\overline{Y} \gets G_{\rm nn}^{\rm gnrt}(Y|\theta_{g})$\;
		$\hat{\boldsymbol{s}} = D_{\rm nn}^{\rm decd}(\overline{Y}|W_{d})$\;
		Updating $W_{e}, W_{d}$ and $\Theta_g$ by computing \eqref{eq:cross entroyl}\;	
		\end{algorithm}

	\textit{c. Training the whole framework}
	
	The last step of AOT is calling Algorithm 1. The pseudo code of training the whole framework is shown in \hyperlink{Algo4-ref}{Algorithm 4}, where $n_{\rm altv}$ is the updating frequency of calling Algorithm 2 and Algorithm 3.  
	
	In the first step of AOT, the GSDSM is updated 1 times by computing $L_G$, and the discriminator is updated 1 time by computing $L_{\rm adv_d}$. In its second step, AEDM and GSDSM are updated simultaneously by computing $L_{\rm CE}$. In the third step, the discriminator is updated 1 time by computing $L_{\rm adv_d}$, and AEDM as well as GSDSM are updated simultaneously by computing $L_{\rm total}$.
	
	\hypertarget{Algo4-ref}{}
	\begin{algorithm}
		\caption{Training the whole framework: AOT.}
		\KwIn{The knowledge base $\mathcal{K}$.}
		\KwOut{The trained network parameters of AEDM $W_{e}$ and $W_{d}$. The trained network parameters of GSDM and discriminator $\boldsymbol{\Theta}_g$ and $\boldsymbol{\Theta}_d$, respectively.}
		\While(\hfill\tcp*[h]{Epoch size $e$}){Epoch $e$ is not zero}{
			\For(\hfill\tcp*[h]{Batch size $B$}){$j=1,...,B$}
			{$\boldsymbol{X} = E_{\rm nn}^{\rm encd}(\boldsymbol{s_j}|W_{e})$\;
				Sending $\boldsymbol{x}$ to the AWGN, Rician fading \leavevmode\newline
				or Rayleigh fading channels\;
				\eIf{$j \  \% \ n_{altv} == 0$}{
				Training GSDSM and discrimintor()\;	\hfill\tcp*[h]{Calling Algorithm 2}	
				
				Training AEDM with GSDSM()\;	\hfill\tcp*[h]{Calling Algorithm 3}	
				}
				{
				Training the whole framework()\;	\hfill\tcp*[h]{Calling Algorithm 1}
				}     
			}
		}
		\end{algorithm}

\section{Experiments}
	In this section, we present some experimental results to discuss the performance of our proposed framework. To verify the effectiveness of Ti-GSC (also denoted as GAN-SC-w/o-CSI in this section), we use BLEU as the performance metric. The two presented training schemes (i.e., JOT and AOT) are also compared and discussed. Moreover, to further compare the SDS performance of the two training schemes, normalized mean squared error (nMSE) is used as the performance index.
\subsection{Experiments Setting}
\subsubsection{Dataset}The dataset used for training and testing phases is European Parliament Proceedings Parallel Corpus \cite{Europarl_2005}, which comprises about 2.2 million sentences composed of 53 million words. We preprocess each sentence into 4 to 30 words and divide it into training sets and testing sets. 
\subsubsection{Neural Network Architecture} 
\begin{figure}[!t]
	\centering
	\begin{minipage}{0.49\textwidth}
		\hspace{-0.56cm}
		\subfloat{\includegraphics[width=1.1\textwidth]{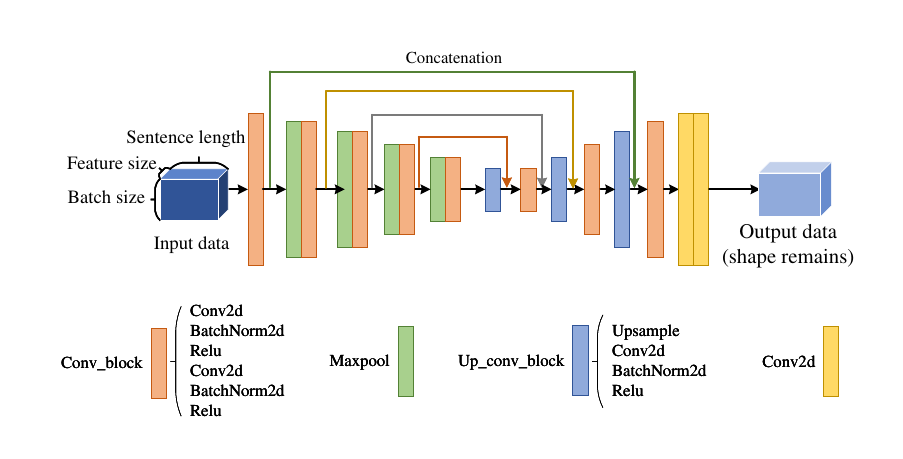}
	}
	\caption{The neural network architecture of GSDSM.}
	\label{Unet_ref}
\end{minipage}
\end{figure}

AEDM consists of multiple transformer layers which is composed of 8 heads as in \cite{xie_deep_2021}. GSDSM aims to learn the distribution of both channel effect and semantic noise over distorted signal, while to eliminate the influence, which emphasizes the ability to perceive semantics and semantic noise. Regarding the above requirement, U-Net \cite{ronneberger_u-net_2015} which is composed of 2 symmetrical networks is adopted as the network architecture of GSDSM. We intend to use both feature concatenation and layer skipping connection of U-Net to preserve the semantic information and also reduce the influence of both channel effect and semantic noise. Therefore, GSDSM is realized by a typical U-net architecture with 5 U-net encoder blocks, 4 U-net decoder blocks and 2 convolutional layers, as illustrated in  Fig. \ref{Unet_ref}. ReLU \cite{glorot_deep_2010} is used as the activation function for AEDM and GSDSM. Softmax is applied for the final output layer of semantic decoder of AEDM. The discriminator contains 4 encoder-layer blocks composed of one convolutional layer, 1 batchnorm layer and leakyrelu \cite{maas_rectier_nodate} activation function, and the last layer of discriminator is a convolutional layer.

\subsubsection{Training and Testing Setting}The initial learning rate for AEDM is set as $10^{-4}$ with $5 \times 10^{-4}$ weight decay, and the learning rate for both $G_{\rm nn}^{\rm gnrt}$ and $D_{\rm nn}^{\rm dscm}$ is $2 \times 10^{-4}$. Adam \cite{Kingma:2014vow} algorithm is adopted for network updating. The range of weight parameter $\lambda$, $\gamma$ and $ w_{k}$ are $[0, +\infty)$, where $k \in \left\{1,2,3 \right\} $. The number of training epoch is set to 60. The testing SNR is set from $0$ dB to $24$ dB, and the interval is $3$ dB. The AWGN channel, Rician fading channel and Rayleigh fading channel are considered in experiments.

\subsubsection{Baselines}
For clarity, in the experiments, our proposed framework for training using the JOT is called \textbf{GAN-SC-w/o-CSI (i.e., Ti-GSC with JOT)}, and it with AOT is called \textbf{GAN-SC-w/o-CSI (i.e., Ti-GSC with AOT)}. We adopt 3 SC frameworks based on ANN, which are the same as our proposed framework architecture but without GSDSM, and 4 conventional frameworks with 16 QAM modulation as baselines.
\begin{itemize}
	\item{\textbf{SC-w/o-CSI (Vanilla SC)}}: In this framework, we remove GSDSM of our proposed framework as the baseline to explore the impact of channel fading on the SC framework.
	\item{\textbf{SC-w-CSI (Perfect CSI)}}: In this framework, we assume that the decoder can get the perfect CSI for channel equalization. The network architecture is the same as our proposed Ti-GSC except for the removal of GSDSM.
	\item{\textbf{SC-w-CSI (Imperfect CSI)}}: Different from SC-w-CSI (Perfect CSI), the imperfect CSI $\boldsymbol{H_e} = \boldsymbol{H} + \boldsymbol{e}$ is considered for decoder owing to outdated channel feedback, where $\boldsymbol{e}$ is the channel estimation error, which is assumed to follow standard circularly symmetric complex Gaussian distribution. In our simulations, the variance $\textit{v}^{2} = 0.002, 0.02, 0.2$.
	
	\item{\textbf{Conventional-w-CSI (Fixed-CC)}}: This framework combines the fixed 5-bit length source coding and the convolutional channel coding \cite{elias1955coding} with perfect CSI for channel equalization.
	\item{\textbf{Conventional-w/o-CSI (Fixed-CC)}}: The framework is identical to the Conventional-w-CSI (Fixed-CC), with the exception that channel equalization is not performed with CSI. 
	\item{\textbf{Conventional-w-CSI (Huffman-CC)}}: We construct the Huffman codebook by computing character frequencies as a dictionary for source coding, and the channel coding is the same as Fixed-CC. The perfect CSI is employed for channel equalization.
	\item{\textbf{Conventional-w/o-CSI (Huffman-CC)}}: The framework remains consistent with Conventional-w-CSI (Huffman-CC), but without using CSI for channel equalization.
	
\end{itemize}
\subsubsection{Performance Metric}
 BLEU \cite{papineni_bleu_2002} is used as the performance index. $P_{k}$ is the modified precision score, and the value of BLEU score is computed by

\begin{equation}
	\label{eq:BLEU}
	BLEU= \rm exp \left(\sum_{k=1}^{K}\frac{1}{n}\log(P_{k})\right) * \rm exp\left({1-\frac{l_{\rm s}}{l_{\hat{s}}}}\right),
\end{equation}
where $l_{\rm s}$ and $l_{\rm \hat{s}}$ are the length of the transmitted text and the length of the decoded text, respectively.

The effect of distortion suppression is evaluated using nMSE, which measures the difference between the transmitted signal and the signal after distortion suppression. The nMSE is defined by,
\begin{equation}
	\label{eq:nMSE}
	nMSE=\frac{\sum_{i=1}^{N}\sum_{j=1}^{M}(x_{ij}-\overline{y}_{ij})^{2}}
	{\sum_{i=1}^{n}\sum_{j=1}^{m}(x_{ij})^{2}},
\end{equation}
where $x_{ij}\in\boldsymbol{x}$ and $\overline{y}_{ij}\in\boldsymbol{y}$ are the signal variables of the $i$-th sentences $j$-th encoded signal and distortion-suppressed  signal, respectively.
\subsection{Comparison of AOT and JOT for Ti-GSC}

\begin{figure}[!t]
	\centering
	\includegraphics[width=3.5in]{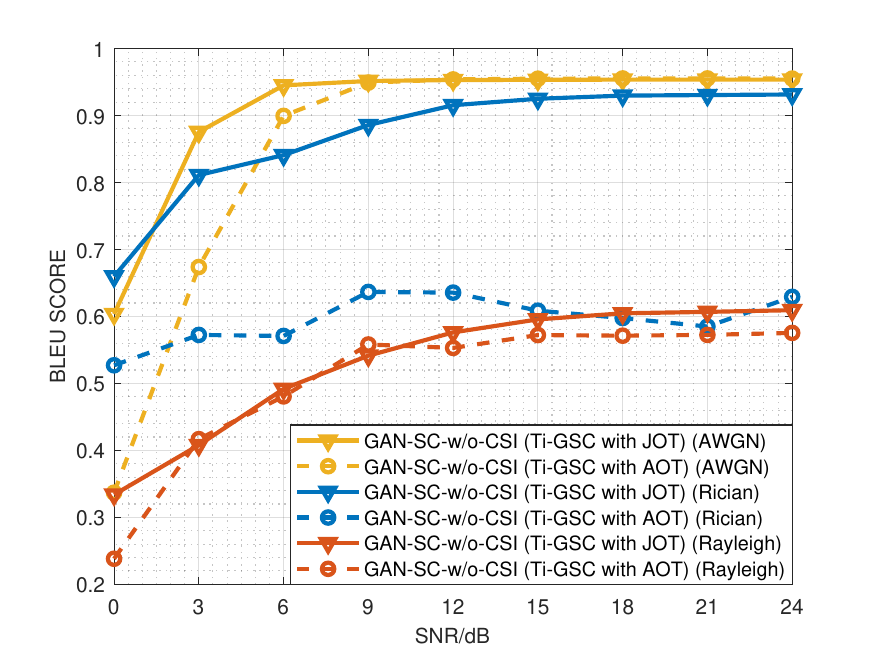}
	\caption{The BLEU scores against SNR of proposed framework through two training schemes.}
	\label{trad_train}
\end{figure}
	In this section, we provide some simulation results to compare JOT and AOT in both AWGN channel and fading channels where both Rician fading and Rayleigh fading are simulated. It can be seen in Fig. \ref{trad_train}, in AWGN channel, the BLEU score of GAN-SC-w/o-CSI (Ti-GSC with AOT) is lower than that of GAN-SC-w/o-CSI (Ti-GSC with JOT) when SNR is below 9 dB, and higher than it when SNR is above 9 dB. It means that GAN-SC-w/o-CSI (Ti-GSC with AOT) reduces the performance of low SNR region to obtain high BLEU scores in high SNR region. Over Rician fading channel, compared with GAN-SC-w/o-CSI (Ti-GSC with JOT), GAN-SC-w/o-CSI (Ti-GSC with AOT) is able to get higher BLEU score, may be caused by the instability of alternative optimization scheme.  Different from the results in Rician fading channel, the results of GAN-SC-w/o-CSI (Ti-GSC with JOT) and  GAN-SC-w/o-CSI (Ti-GSC with AOT) are comparable in Rayleigh fading channel, and both achieving lower BLEU scores. The reason is that the severe signal distortion caused by the Rayleigh channel cannot be well compensated for GAN.

	We also evaluate the GSDSM by nMSE for two training schemes. As shown in Fig. 7 (a) and Fig. 7 (b), both schemes tend to converge with the increase of epoch, and the nMSE achieved in Rayleigh fading channel is higher than in Rician fading channel. For different schemes, the signal recovery ability varies significantly. For GAN-SC-w/o-CSI (Ti-GSC with AOT), the nMSE in Rician and Rayleigh fading channel converges to very lower values compared with GAN-SC-w/o-CSI (Ti-GSC with JOT). However, the results shown in Fig. 7 are opposite the performance achieved as shown in Fig. \ref{trad_train}. That is, regarding GAN-SC-w/o-CSI (Ti-GSC with JOT), although the nMSE is inferior to GAN-SC-w/o-CSI (Ti-GSC with AOT), the result in Fig. \ref{trad_train} shows the good performance of joint optimization. It means that AOT may lead to local optima and unstable training for our proposed Ti-GSC. Note that, as GAN-SC-w/o-CSI (Ti-GSC with AOT) is inferior to GAN-SC-w/o-CSI (Ti-GSC with JOT) after experiments, we only discuss GAN-SC-w/o-CSI (Ti-GSC with JOT) for subsequent experiments. 

\begin{figure}[htbp] 
		\centering
		\captionsetup[subfloat]{justification=centering}
		\begin{minipage}{0.5\textwidth}
		\subfloat[The nMSE of GAN-SC-w/o-CSI (Ti-GSC with JOT) \\ in  Rician fading channle and Rayleigh fading channel.]
		{\includegraphics[width=3.5in]{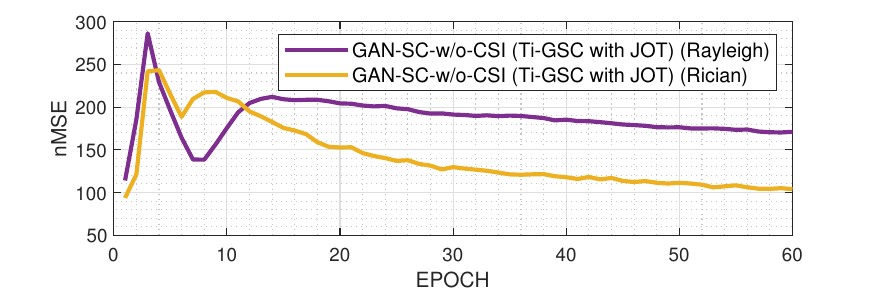}
		\label{nmse_a}}
	\end{minipage}
	\begin{minipage}{0.5\textwidth} 
		\centering
		\subfloat[The nMSE of GAN-SC-w/o-CSI (Ti-GSC with AOT) \\ in  Rician fading channle and Rayleigh fading channel.]{\includegraphics[width=3.5in]{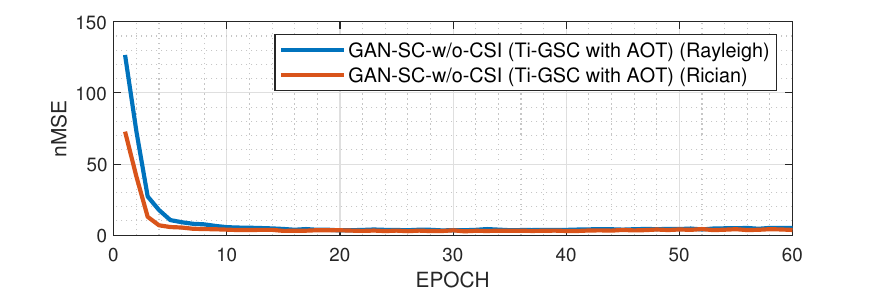}
			\label{nmse_o}}	
		\caption{The nMSE of two kinds of training schemes in Rician fading channle and Rayleigh fading channel.}	
	\end{minipage}

	\label{nMSE}
\end{figure}

\hypertarget{4B_ref}{\subsection{Comparison of Ti-GSC with Other Communication Frameworks without Considering Channel Estimation and Channel Equalization Time Consumption.
}}
\begin{figure}[!t]
	\centering
	\includegraphics[width=3.5in]{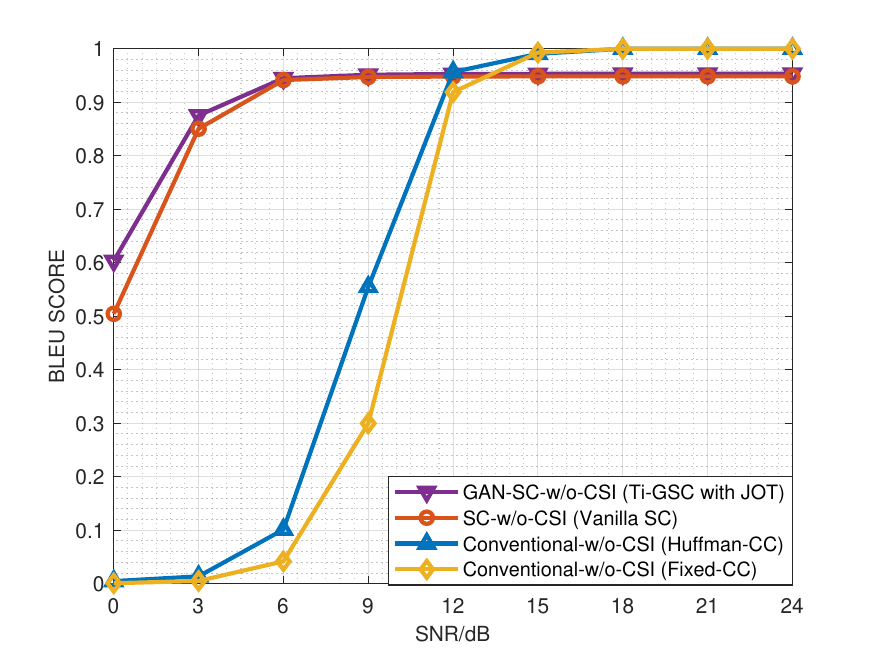}
	\caption{The BLEU scores against SNR of proposed framework and baselines in AWGN channel.}
	\label{trad_aw}
\end{figure}
\begin{figure*}[!t]
	\centering
	\subfloat{\includegraphics[width=5.5in]{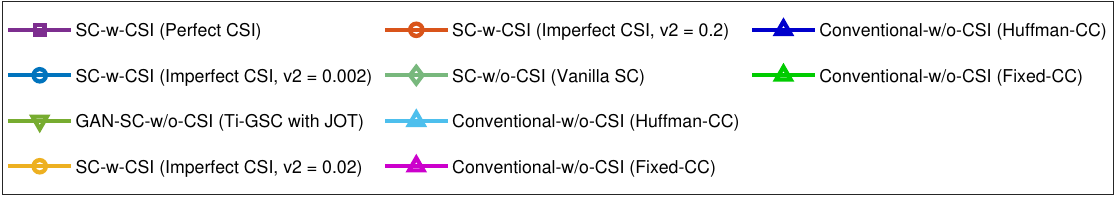}%
		}
	\vspace{-0.35cm}
	\setcounter{subfigure}{0} 
	\subfloat[The BLEU score in Rician fading channel]{\includegraphics[width=3.5in]{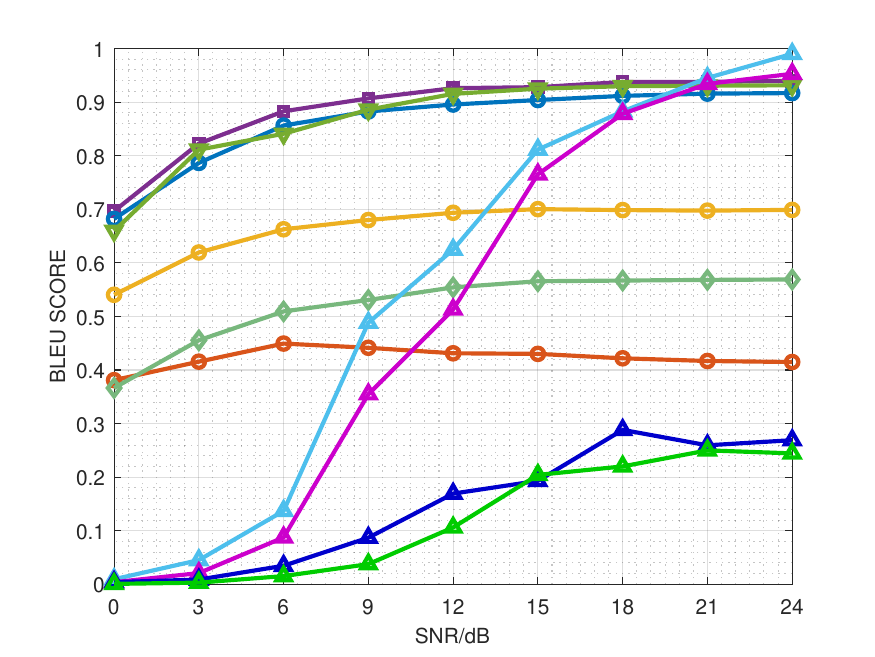}%
		\label{trad_ri}}
	\hfil
	\subfloat[The BLEU score in Rayliegh fading channel]{\includegraphics[width=3.5in]{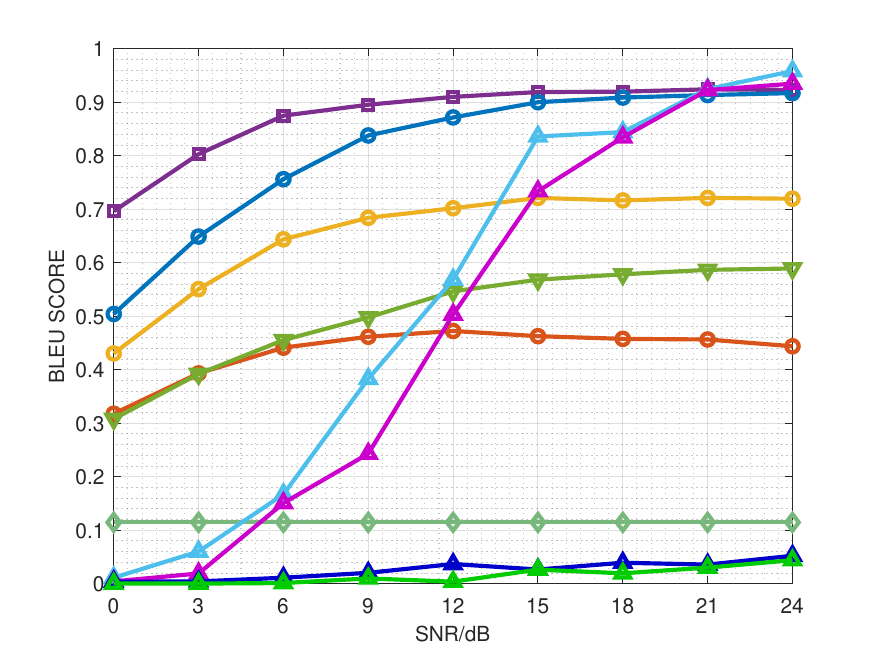}%
		\label{trad_ra}}
	\caption{The BLEU scores against SNRs of proposed framework and baselines in Rician fading channle and Rayleigh fading channel.}
	\label{fig_sim}
\end{figure*}

	In this subsection, we compare our proposed Ti-GSC with ANN-based SC frameworks with CSI and that without CSI. Moreover, conventional frameworks are also simulated for comparison. In Fig. \ref{trad_aw}, we compare the proposed GAN-SC-w/o-CSI (Ti-GSC with JOT) with SC-w/o-CSI (Vanilla SC) and 2 Conventional-w/o-CSI ones in AWGN channels. In Fig. 9 (a) and Fig. 9 (b), we compare the proposed GAN-SC-w/o-CSI (Ti-GSC with JOT) with SC-w/o-CSI (Vanilla SC), SC-w-CSI (Perfect CSI), SC-w-CSI (Imperfect CSI), 2 Conventional-w-CSI and 2 Conventional-w/o-CSI frameworks in both Rician and Rayleigh fading channels.

	It is noted that as SC-w-CSI (Perfect CSI), SC-w-CSI (Imperfect CSI) and 2 Conventional-w-CSI frameworks involve time-consuming processes for conducting channel estimation and channel equalization, but none-CSI frameworks does not involve. In this subsection, we compare our Ti-GSC with baselines by neglecting time consumption of these processes, which means SC-w-CSI (Perfect CSI), SC-w-CSI (Imperfect CSI) and 2 Conventional-w-CSI frameworks are compared under the ideal performance. 
	
	Fig. \ref{trad_aw} reveals that in the low SNR region (0 dB - 9dB) in AWGN channel, the proposed framework and SC-w/o-CSI (Vanilla SC) outperform traditional communication ones. While due to the performance loss caused by the limitation of gradient descent algorithm, SC frameworks based on ANN are slightly weaker than traditional ones in the SNR region above $12$ dB. 
	
	Fig. 9 (a) and Fig. 9 (b) show the BLEU score of different frameworks against SNR in Rician fading channel and Rayleigh fading channel, respectively. As shown in Fig. 9 (a), GAN-SC-w/o-CSI (Ti-GSC with JOT) outperforms baselines except for SC-w-CSI (Perfect CSI), SC-w-CSI (Imperfect CSI, $v^2$= 0.002) below 9 dB and two conventional-w-CSI frameworks above 21 dB. Because SC framework with perfect CSI (SC-w-CSI (Perfect CSI)) is not affected by channel stochasticity of fading channel.  SC framework with imperfect CSI but only slight CSI error (SC-w-CSI (Imperfect CSI, $v^2$ = 0.002)) is also insensitive to the negative effects caused by channel stochasticity. For proposed GAN-SC-w/o-CSI (Ti-GSC with JOT), GSDSM is trained to recover the signal in syntactic dimension and semantic dimension, thus, the input of semantic decoder can be calibrated to get better performance. For SC-w-CSI (Imperfect CSI, $v^2$ = 0.02) and SC-w-CSI (Imperfect CSI, $v^2$ = 0.2), signal distortion deteriorates as error variance increases, leading to a significant decrease in the performance of both frameworks. For SC-w/o-CSI (vanilla), the ability to suppress distortion is limited to syntactic dimension, resulting in lower BLEU scores. For four conventional frameworks, Huffman-cc coding exceeds 5-bit-cc fixed-length coding due to better source coding efficiency. Meanwhile, due to channel equalization, the signal distortion is relatively small, making conventional-w-CSI frameworks superior to the conventional-w/o-CSI ones.
	
	As shown in Fig. 9 (b). SC-w/o-CSI (vanilla) occurs a recovery collapse in Rayleigh fading channel, which may be due to Rayleigh fading channel models a worse channel condition than that of Rician fading channel, and severe signal distortion leading to network learning collapse. The trends of four conventional frameworks are similar to that in Rician fading channel. It is also observed that, the BLEU score of SC-w-CSI (Imperfect CSI, $v^2$ = 0.2) in both Rician and Rayleigh channels increase initially with the increase of SNR, followed by a decreasing trend. Perhaps it is because the substantial noise present in CSI estimation appears to have a dual impact: not only does it degrade performance, but it also stands out as a primary factor adversely influencing the learning process of both the encoder and decoder, and this influence surpasses that of SNR. That is, an increase in SNR does not always increase BLEU scores of SC-w-CSI (Imperfect CSI, $v^2$ = 0.2).

\subsection{Comparison of Ti-GSC with Other Communication Frameworks with Considering Channel Estimation and Channel Equalization Time Consumption.}

\begin{figure*}[htbp]   
	\centering
	\captionsetup[subfloat]{justification=centering}
	
	\begin{minipage}{1\textwidth} 
		\centering	
		\subfloat[The BLEU score against preprocessing time consumption in Rician fiading channel\\ in SNR = 18 dB.]{\includegraphics[width=0.33\textwidth]{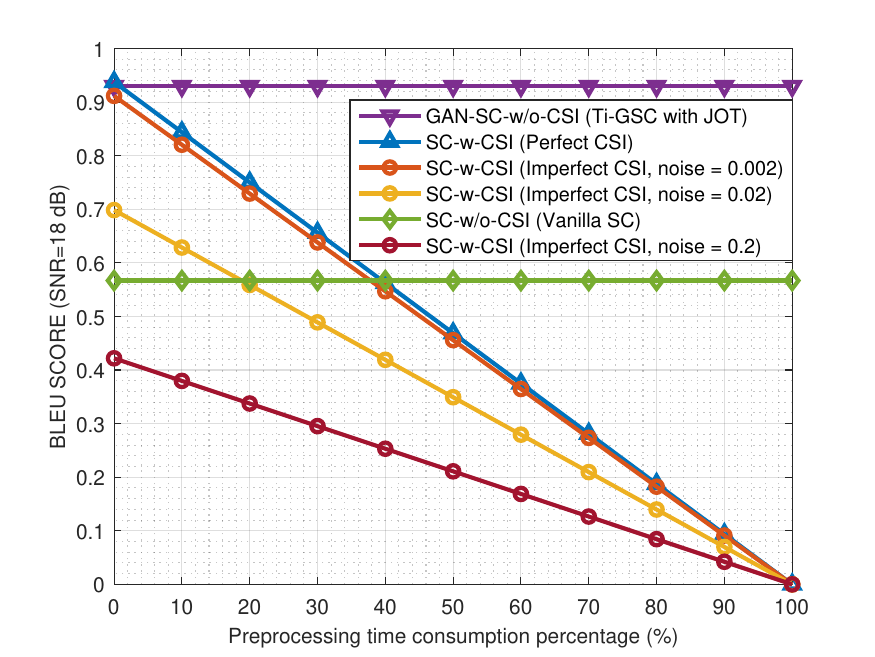}
			\label{time_ri_18}}	
		\subfloat[The BLEU score against preprocessing time consumption in Rician fiading channel\\ in SNR = 9 dB.]{\includegraphics[width=0.33\textwidth]{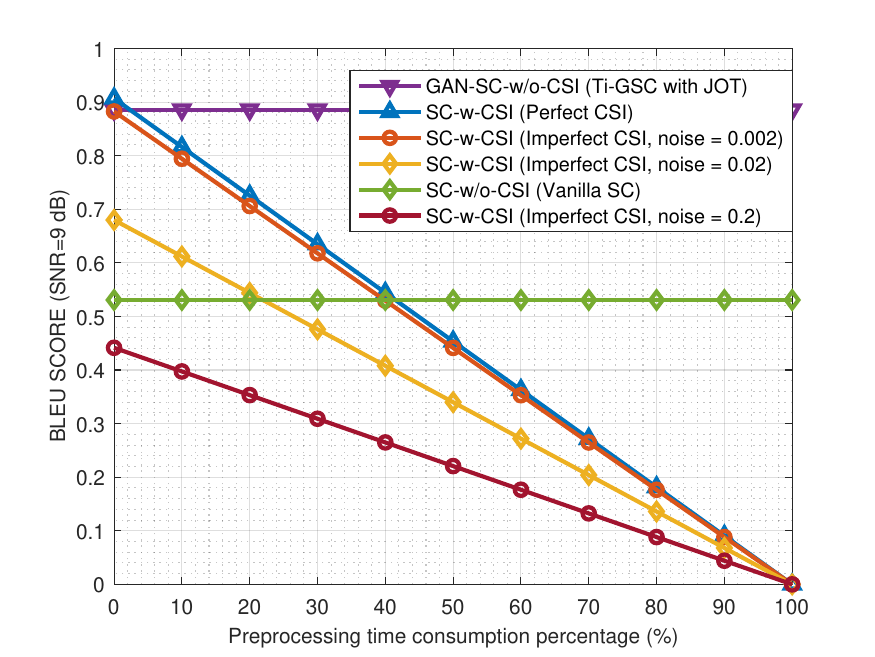}
	\label{time_ri_9}}	
		\subfloat[The BLEU score against preprocessing time consumption in Rician fiading channel\\ in SNR = 0 dB.]{\includegraphics[width=0.33\textwidth]{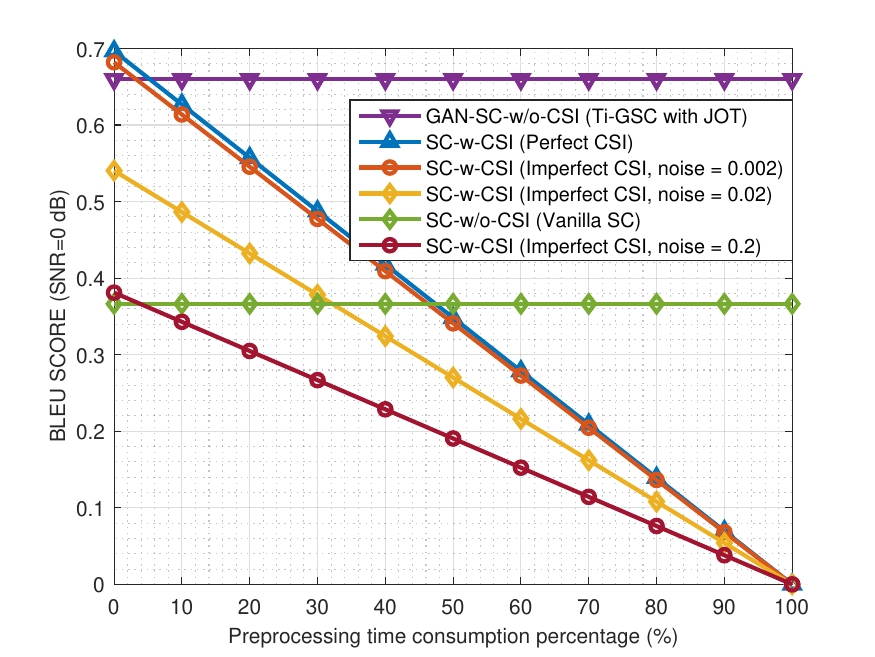}
	\label{time_ri_0}}	
	\end{minipage}
	\begin{minipage}{1\textwidth} 
		\centering	

		\subfloat[The BLEU score against preprocessing time consumption in Rayleigh fiading channel\\ in SNR  = 18 dB.]{\includegraphics[width=0.33\textwidth]{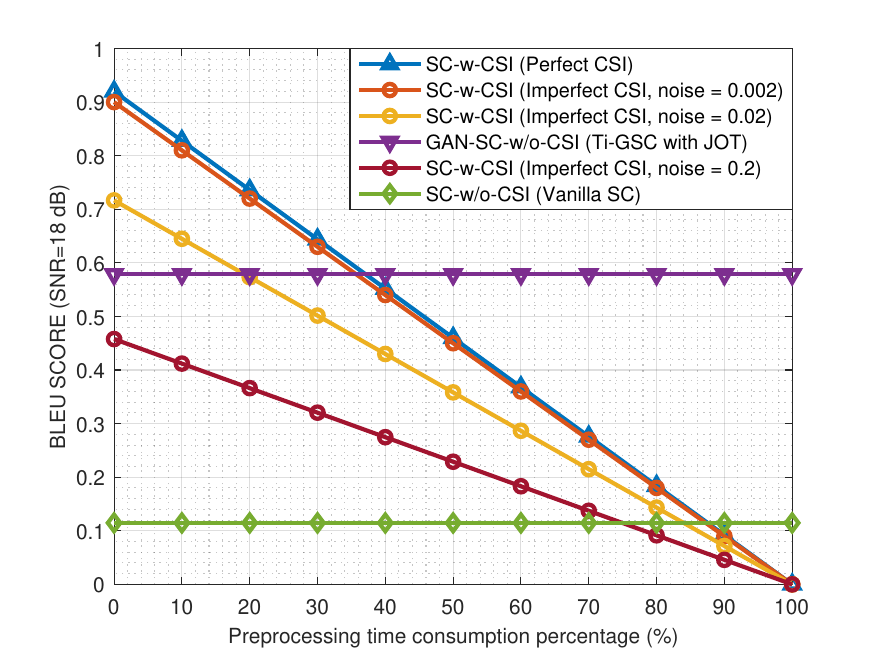}
	\label{time_ra_18}}	
		\subfloat[The BLEU score against preprocessing time consumption in Rayleigh fiading channel\\  in SNR  = 9 dB.]{\includegraphics[width=0.33\textwidth]{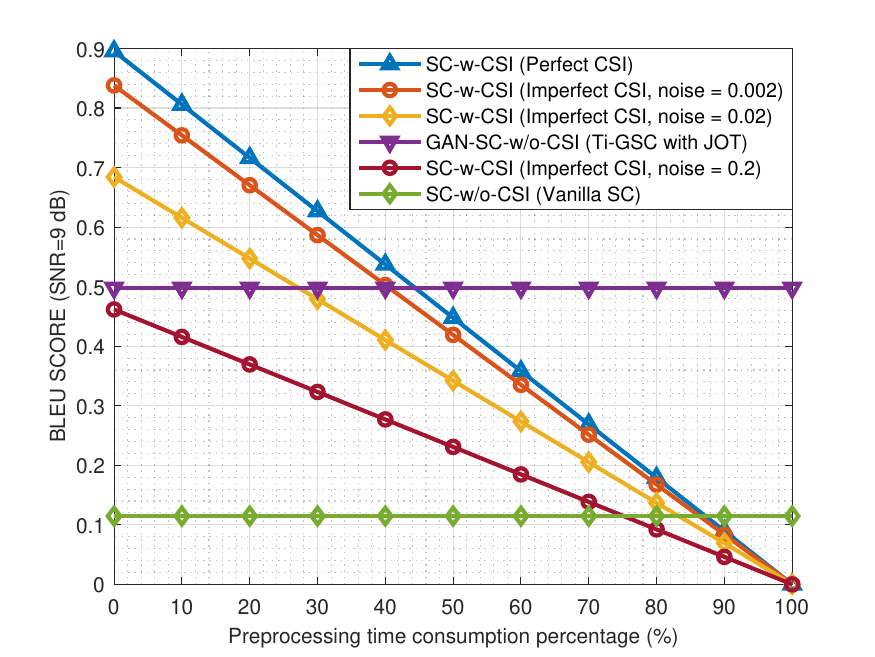}
			\label{time_ra_9}}	
		\subfloat[The BLEU score against preprocessing time consumption in Rayleigh fiading channel\\ in SNR  = 0 dB.]{\includegraphics[width=0.33\textwidth]{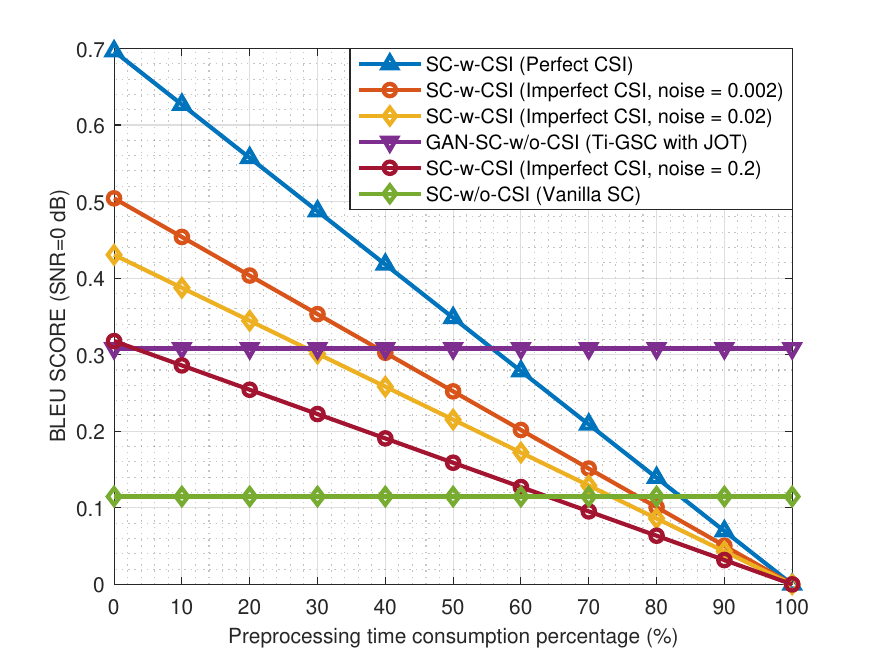}
	\label{time_ra_0}}	
	\end{minipage}
	\begin{minipage}{1\textwidth} 
		\centering

		\caption{The BLEU scores against preprocessing time consumption in Rayleigh fading channle and Rayleigh fading channel.}
	\end{minipage}
	
\end{figure*}

To achieve a more fair comparison, time consumption is taken into account in this subsection. We employ the time proportion method to characterize the performance degradation resulting from the time consumption of channel estimation and channel equalization, which means the actual performance of frameworks with CSI needs to be discounted based on the time consumed for channel estimation and channel equalization. For simplicity, the time consumption of channel estimation and channel equalization are collectively referred to as preprocessing time consumption. For instance, if the preprocessing consumption time is 10 $\%$ of a transmission time slot, the real performance needs to be multiplied by 0.9. Fig. 10 shows the real BLEU scores against preprocessing time consumption of ANN-based frameworks in Rician fading channel and Rayleigh fading channel in SNR=18 dB, 9 dB and 0 dB. 

As shown in Fig. 10, the trend of GAN-SC-w/o-CSI (Ti-GSC with JOT) and SC-w/o-CSI (Vanilla SC) remains unchanged as CSI is not required. As shown in Fig. 10 (a), Fig. 10 (b) and Fig. 10 (c), when the percentage of preprocessing time consumption is above about 5 $\%$, GAN-SC-w/o-CSI (Ti-GSC with JOT) outperforms baselines and the gain increases with the increase of preprocessing time consumption in SNR = 0,9 and 18 dB. When half of the time is used for preprocessing, SC-w/o-CSI (Vanilla SC) outperforms others except for our proposed one. As shown in Fig. 10 (d), Fig. 10 (e) and Fig. 10 (f), the horizontal coordinates where GAN-SC-w/o-CSI (Ti-GSC with JOT) intersects with SC-w/o-CSI (Perfect CSI) are approximately 38 $\%$, 45 $\%$ and 56 $\%$, respectively, which are much greater than the percentages required in Rician fading channel. Because proposed non-CSI framework has limited performance under severe channel fading such as Rayleigh fading channel.

\subsection{Ablation Experiments}

As shown in equation (12), the designed total loss function $L_{\rm total}$ includes four terms: $L_{\rm CE}$, $L_{\rm adv\_g}$, $ L_{\rm sytc}$and $L_{\rm dstr}$. To reveal the different influences of newly added terms ($L_{\rm sytc}$ and $L_{\rm dstr}$) and adversarial learning ($L_{\rm adv\_g}$) on proposed Ti-GSC, we perform ablation experiments by using control variables. Specifically, we set $w_{2} = 0$ for training without adversarial learning, $\lambda = 0$ for training without syntactic learning and $\gamma = 0$ for training without semantic learning.

As shown in Fig. 11 (a), the most significant term of loss is syntactic learning, as the BLEU score of training without syntactic learning is lowest in Rician fading channel. The performance degradation of training without semantic learning is less than training without syntactic learning, and the result is consistent with semantic distortion may not occur even when syntactic distortion occurs. Training without adversarial learning has a relatively small impact on performance, as it results in a relatively small performance loss. Compared training without adversarial learning with training without syntactic learning and training without semantic learning, it indicates that GSDSM emphasizes the importance of recovering ability over generating ability in Rician fading channel. Thus, training without adversarial learning achieves a better BLEU score than training without syntactic learning and semantic learning.

Figure 11 (b) shows different results from Rician fading, that is, adversarial learning occupies a more important position, syntactic learning contributes to score enhancement in region of high SNR (above 12 dB), whereas semantic learning leads to score improvement in low SNR region (below 12 dB) in Rayleigh fading channel. The experiment of training without adversarial learning shows that the ability of complete self-supervised learning with restricted generation capability cannot compensate for all the distortions well. Additionally, the input of GSDSM has no prior information which leads to higher requirements for signal remodeling capability. In other words, without generative capacity, ANN-based models lack efficacy in handling more intricate issues. It is also conducted to clarify why we use a GAN-based network to suppress the distortion, and particularly noticeable in Fig. 11 (b).

\begin{figure*}[!t]
	\centering
	\subfloat[The BLEU score in Rician fading channel]{\includegraphics[width=3.5in]{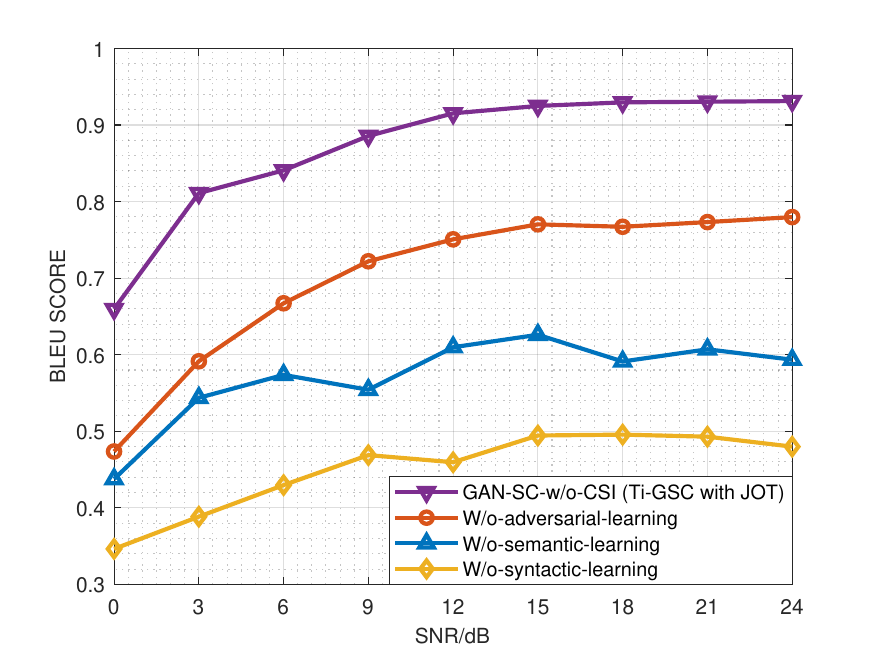}%
		\label{abla_ri}}
	\hfil
	\subfloat[The BLEU score in Rayleigh fading channel]{\includegraphics[width=3.5in]{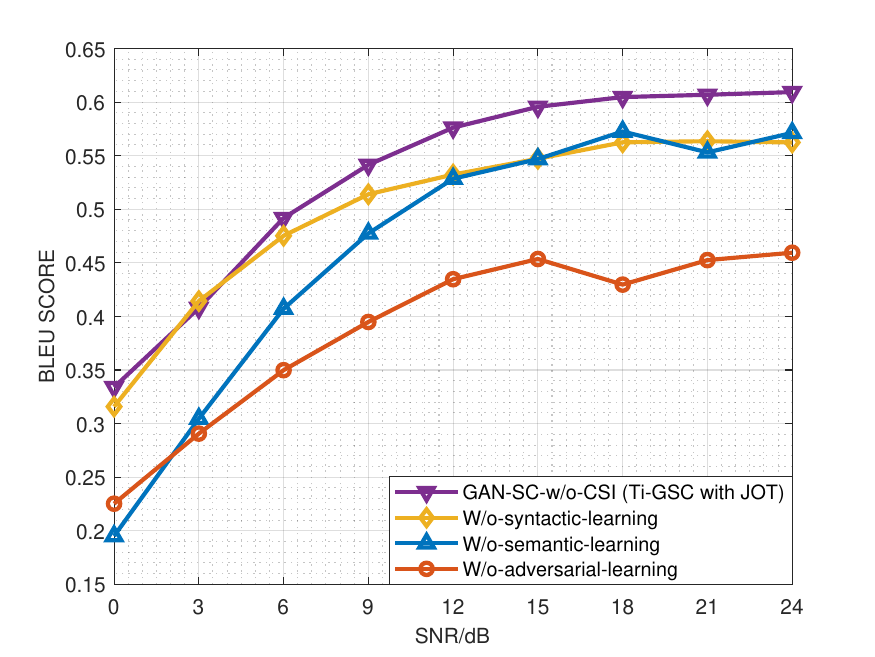}%
		\label{abla_ra}}
	\caption{The BLEU scores against SNR for ablation experiment in Rician fading channle and Rayleigh fading channel.}
\end{figure*}

\hypertarget{4E-ref}{\subsection{ Visual Comparison Examples}}
\begin{figure}[!t]
	\centering
	\includegraphics[width=3.5in]{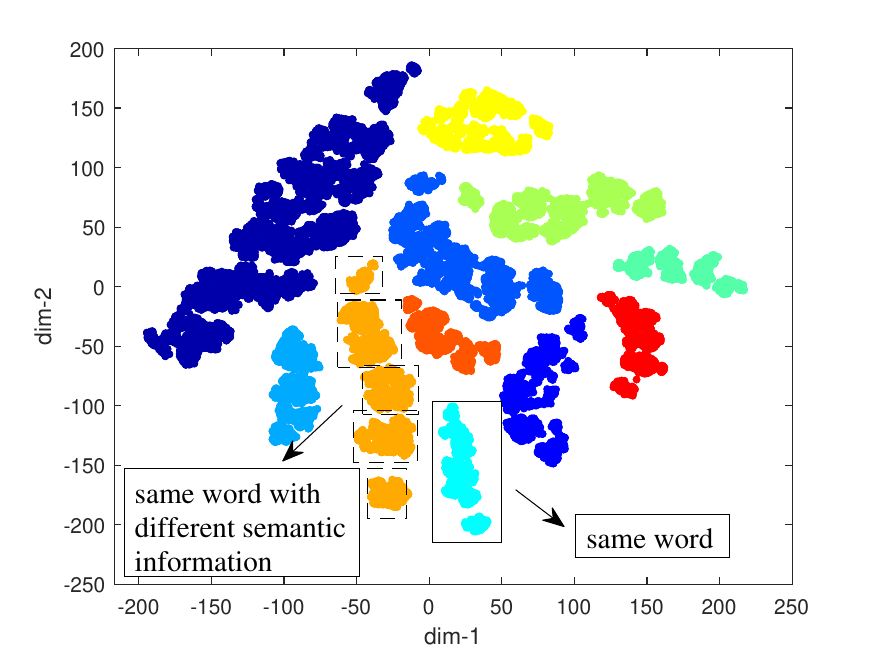}
	\caption{The visualization of encoded signal at the source.}
	\label{trans_sig}
\end{figure}

\begin{figure}[htbp]   
	\centering
	\captionsetup[subfloat]{justification=centering}
	\begin{minipage}{0.5\textwidth} 
	\centering	
	\subfloat[The visualization of received signal at the destination \\ in SNR = 24 dB.]{\includegraphics[width=0.48\textwidth]{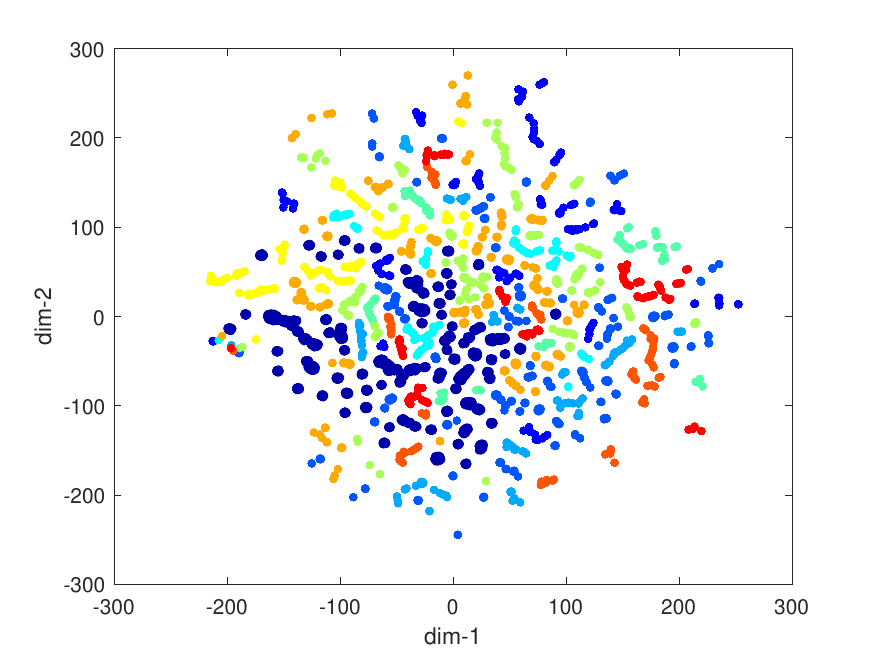}
		\label{vil_rs24}}	
	\subfloat[The visualization of distortion-suppressed signal \\ in SNR  = 24 dB.]{\includegraphics[width=0.48\textwidth]{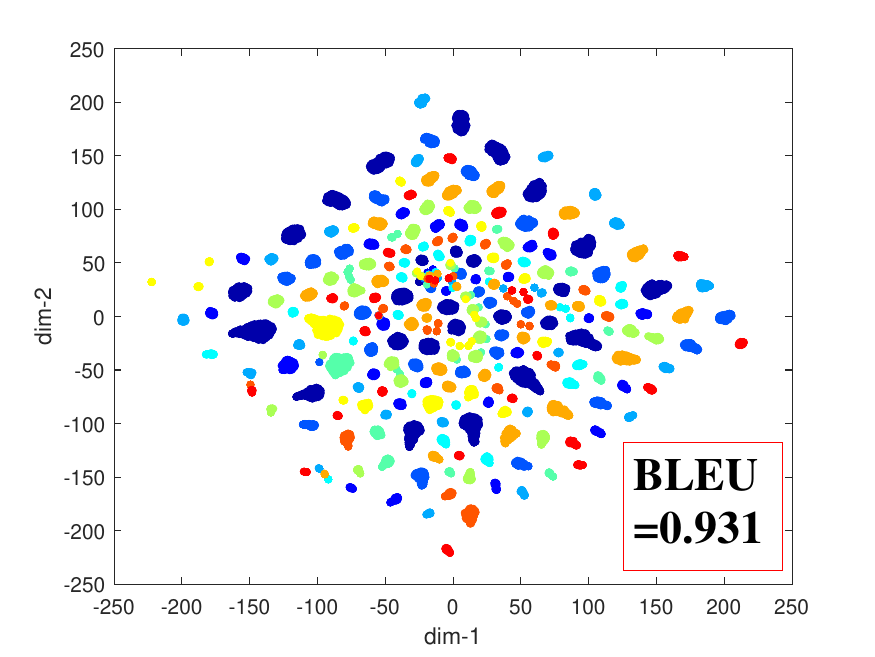}
		\label{vil_sds24}}	
	\end{minipage}

	\begin{minipage}{0.5\textwidth} 
		\centering	
		\subfloat[The visualization of received signal at the destination \\ in SNR = 12 dB.]{\includegraphics[width=0.48\textwidth]{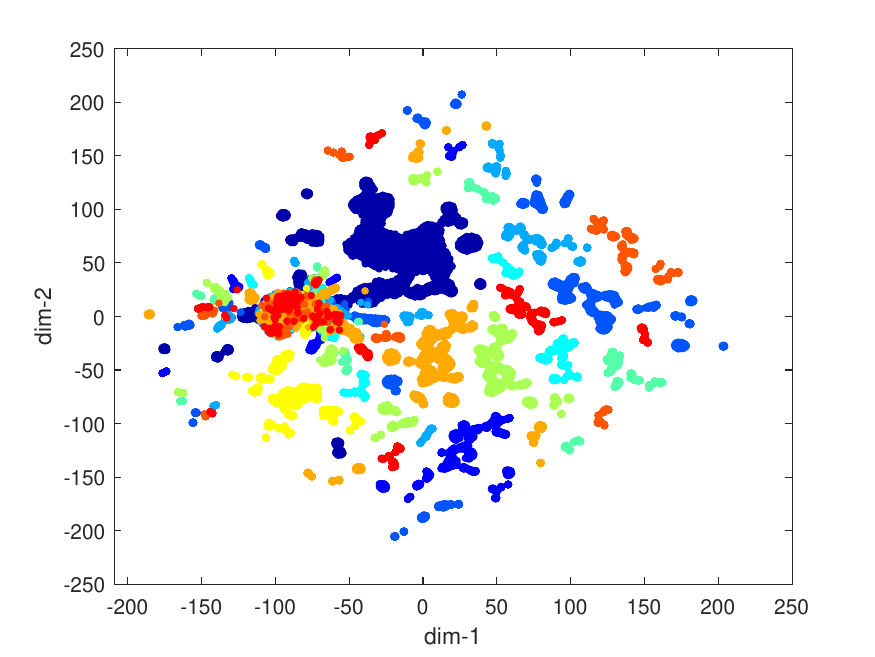}
			\label{vil_rs12}}	
		\subfloat[The visualization of distortion-suppressed signal \\ in SNR  = 12 dB.]{\includegraphics[width=0.48\textwidth]{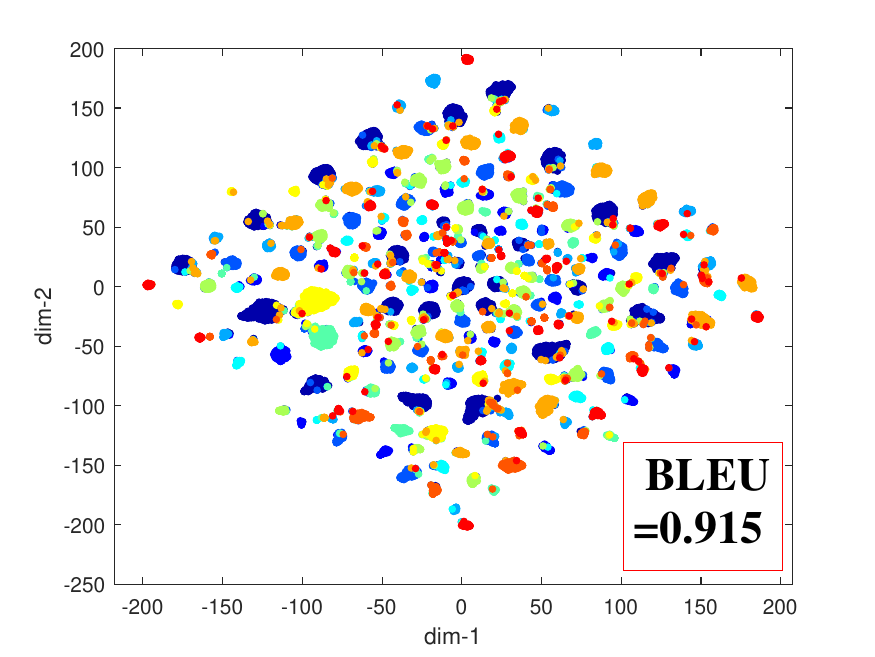}
			\label{vil_sds12}}	
	\end{minipage}

	\begin{minipage}{0.5\textwidth} 
		\centering
		\subfloat[The visualization of received signal at the destination \\ in SNR = 0 dB.]{\includegraphics[width=0.48\textwidth]{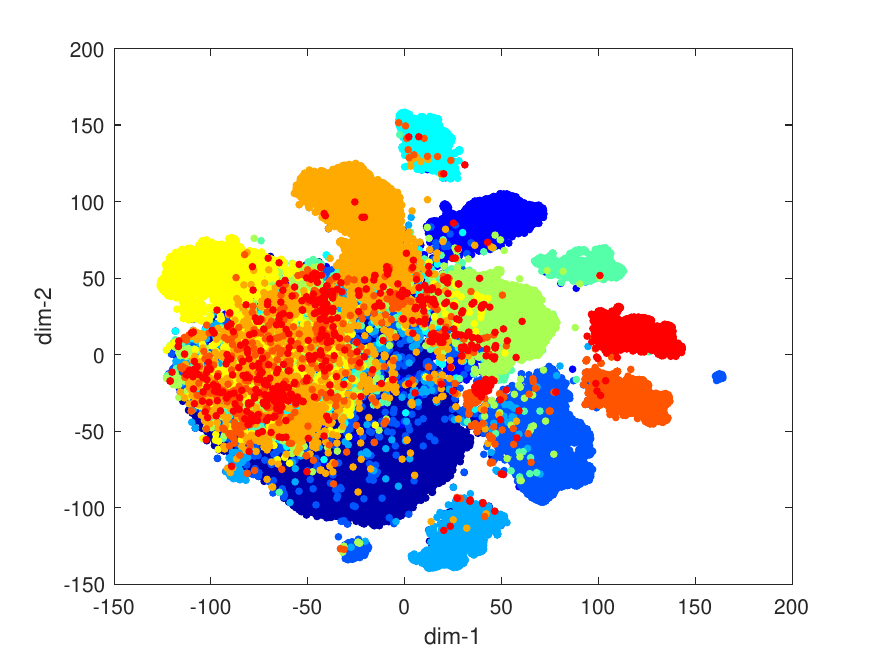}
			\label{vil_rs0}}		
		\subfloat[The visualization of distortion-suppressed signal \\ in SNR  = 0 dB.]{\includegraphics[width=0.48\textwidth]{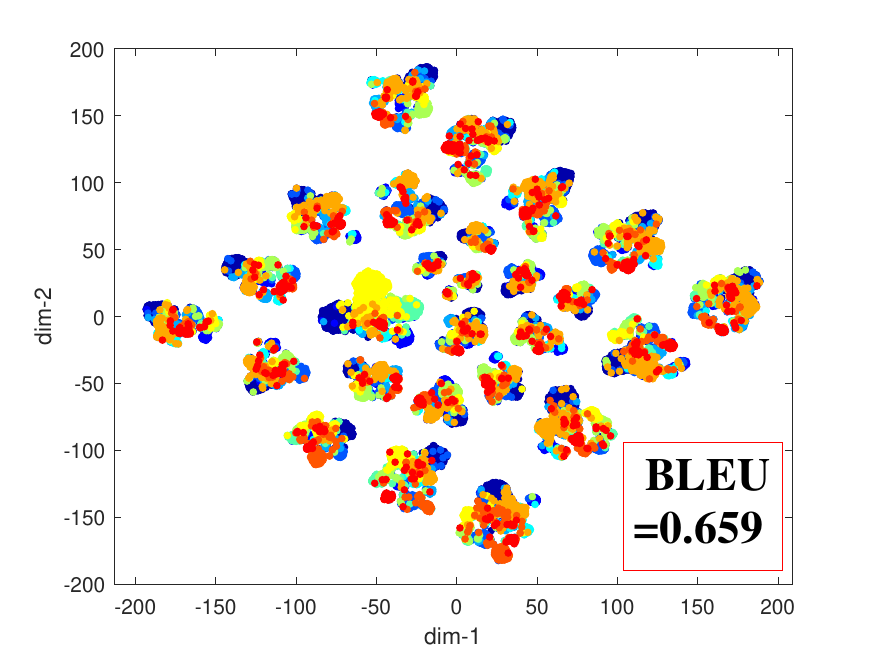}
			\label{vil_sds0}}
		\caption{Fig. (a) (c) and (e) are the visualizations of received signals in SNR = 24 dB,  12dB, and 0dB, respectively; Fig. (b) (d) and (f) are the visualizations of distortion-suppressed signal in SNR = 24 dB,  12dB, and 0dB, respectively.}
		\label{tSNE}		
	\end{minipage}
\end{figure}
In previous subsection, we discuss the performance of GAN-SC-w/o-CSI (Ti-GSC with JOT) in terms of BLEU. In order to offer a more visually comprehensible illustration of the impact of encoding sentences with AEDM and performing SDS using GSDSM on received signals, we provide a visual example. To visual the high-dimensional signal in a simply way, we utilize the t-SNE \cite{laurens_visualizing_2008} algorithm to visualize the spatial distribution of transmitted signal, the received signal and the distortion-suppressed signal of the proposed GAN-SC-w/o-CSI (Ti-GSC with JOT) in Rician fading channel. In order to make the visual results more intuitive, we sorted the words according to their frequency and selected 11 words with the highest frequency.

As shown in Fig. \ref{trans_sig}, it can be seen from the spatial distribution of visual presentation that different words encoded by encoder present different clusters. At the same time, because even the same word performs different functions in different sentences, its semantic mapping presents many small clusters in the identical cluster. In other words, syntactic information distinguishes different words, and semantic information distinguishes different meanings of the same word. 

It is observed from Fig. 13 (a), Fig. 13 (c) and Fig. 13 (e) that the distribution of received signals presents increasingly disorder as the channel SNR decreases. As shown in Fig. 13 (e), only large clusters remain but with overlapping ,which means the semantic information is lost. Fig. 13 (a) displays that both syntactic and semantic information are preserved in high SNR, which just is the reason that a high BLEU score is obtained in high SNR regime by Ti-GSC. 

Figure. 13 (b) shows a special distribution result which is different from transmitted signals as shown in Fig. \ref{trans_sig}. The reason is that the encoder encodes the signal must concern the tradeoff between reliable transmission (syntactic information) and more detailed recovery (semantic information) based on reliable transmission. On the contrary, the signal generated by the GSDSM is transmitted to the decoder without propagation distortion. To achieve optimal performance, the joint optimization technique balances the distribution of the encoder and adjusts the GSDSM towards a semantic-oriented distribution for semantic decoder decoding. 

By comparing Fig. 13 (b), Fig. 13 (d) and Fig. 13 (f), it is evident that the distributions of the distortion-suppressed signals exhibit similarity under different SNR conditions, although the received signal encounters severe disorder in low SNR. The observations of distortion-suppressed signal and the achieving BLEU scores substantiate the effectiveness of GSDSM, which proves that the proposed algorithm does not require additional prior information as input and also demonstrate the robustness of our proposed GAN-SC-w/o-CSI (Ti-GSC with JOT) framework across varying SNR conditions. Consequently, our proposed framework saves on communication costs and enhances overall communication efficiency.

\section{Conclusion}
In this study, we proposed a non-CSI SC framework for text input, named Ti-GSC, which is composed of a GSDSM and an AEDM. We designed and utilized the GSDSM to collaborate with the AEDM to suppress the signal distortion in syntactic and semantic dimensions. To train the framework effectively, we designed a distortion measurement loss function and a differentiable semantic distortion measurement method. Moreover, two training schemes, (i.e., JOT and AOT) were designed for Ti-GSC. Experiments showed JOT is effective for Ti-GSC. In addition, compared with conventional frameworks without CSI, our proposed framework Ti-GSC is more effective. Besides, the BLEU score was improved by approximately 40$\%$ and 62$\%$ with the integration of GSDSM into the presented Ti-GSC framework, in contrast to framework that GSDSM is not utilized in Rician fading channel and Rayleigh fading channel, respectively. Ablation experiments demonstrated the significance of each term within the loss function of GSDSM with respect to the performance of Ti-GSC. Additionally, it is worth noting that the influence of syntactic learning on GSDSM is observed to be the most important in Rician fading channel, surpassing that of semantic learning and adversarial learning. While, in Rayleigh fading, adversarial learning takes precedence as the most crucial factor.

\bibliography{MaoJin_BJTU_arxiv_submitted}

\begin{thebibliography}{10}
\providecommand{\url}[1]{#1}
\csname url@samestyle\endcsname
\providecommand{\newblock}{\relax}
\providecommand{\bibinfo}[2]{#2}
\providecommand{\BIBentrySTDinterwordspacing}{\spaceskip=0pt\relax}
\providecommand{\BIBentryALTinterwordstretchfactor}{4}
\providecommand{\BIBentryALTinterwordspacing}{\spaceskip=\fontdimen2\font plus
\BIBentryALTinterwordstretchfactor\fontdimen3\font minus
  \fontdimen4\font\relax}
\providecommand{\BIBforeignlanguage}[2]{{%
\expandafter\ifx\csname l@#1\endcsname\relax
\typeout{** WARNING: IEEEtran.bst: No hyphenation pattern has been}%
\typeout{** loaded for the language `#1'. Using the pattern for}%
\typeout{** the default language instead.}%
\else
\language=\csname l@#1\endcsname
\fi
#2}}
\providecommand{\BIBdecl}{\relax}
\BIBdecl

\bibitem{luan20226g}
N.~Luan, K.~Xiong, Y.~Zhang, R.~He, G.~Qu, and B.~Ai, ``6g: Typical
  applications, key technologies and challenges,'' \emph{Chin. J. Internet
  Things}, vol.~6, no.~1, pp. 29--43, 2022.

\bibitem{6704826}
K.~Xiong \emph{et~al.}, ``Optimal cooperative beamforming design for mimo
  decode-and-forward relay channels,'' \emph{IEEE Trans. Signal Process.},
  vol.~62, no.~6, pp. 1476--1489, 2014.

\bibitem{9140412}
J.~Liu \emph{et~al.}, ``Max-min energy balance in wireless-powered hierarchical
  fog-cloud computing networks,'' \emph{IEEE Trans. Wireless Commun.}, vol.~19,
  no.~11, pp. 7064--7080, 2020.

\bibitem{8334613}
Y.~Lu \emph{et~al.}, ``Coordinated beamforming with artificial noise for secure
  swipt under non-linear eh model: Centralized and distributed designs,''
  \emph{IEEE J. Select. Areas Commun.}, vol.~36, no.~7, pp. 1544--1563, 2018.

\bibitem{10032267}
R.~Zhang, K.~Xiong \emph{et~al.}, ``Energy efficiency maximization in
  ris-assisted swipt networks with rsma: A ppo-based approach,'' \emph{IEEE J.
  Sel. Areas Commun.}, vol.~41, no.~5, pp. 1413--1430, 2023.

\bibitem{10233705}
C.~Meng \emph{et~al.}, ``Sum-rate maximization in star-ris assisted rsma
  networks: A ppo-based algorithm,'' \emph{IEEE Internet Things J.}, pp. 1--1,
  2023.

\bibitem{10163978}
H.~Zheng, K.~Xiong \emph{et~al.}, ``Maximizing age-energy efficiency in
  wireless powered industrial ioe networks: A dual-layer dqn-based approach,''
  \emph{IEEE Trans. Wireless Commun.}, pp. 1--1, 2023.

\bibitem{dang_what_2020}
S.~Dang, O.~Amin, B.~Shihada, and M.-S. Alouini, ``\BIBforeignlanguage{en}{What
  should {6G} be?}'' \emph{\BIBforeignlanguage{en}{Nat Electron}}, vol.~3,
  no.~1, pp. 20--29, Jan. 2020.

\bibitem{wang_survey_2023}
\BIBentryALTinterwordspacing
Y.~Wang, Z.~Su, S.~Guo, M.~Dai, T.~H. Luan, and Y.~Liu, ``A {Survey} on
  {Digital} {Twins}: {Architecture}, {Enabling} {Technologies}, {Security} and
  {Privacy}, and {Future} {Prospects},'' Jan. 2023, arXiv:2301.13350. [Online].
  Available: \url{http://arxiv.org/abs/2301.13350}
\BIBentrySTDinterwordspacing

\bibitem{zhang_towards_2022}
\BIBentryALTinterwordspacing
S.~Zhang, W.~Y.~B. Lim, W.~C. Ng, Z.~Xiong, D.~Niyato, X.~S. Shen, and C.~Miao,
  ``Towards {Green} {Metaverse} {Networking} {Technologies}, {Advancements} and
  {Future} {Directions},'' Nov. 2022, arXiv:2211.03057. [Online]. Available:
  \url{http://arxiv.org/abs/2211.03057}
\BIBentrySTDinterwordspacing

\bibitem{weaver_weaver_nodate}
W.~Weaver, ``\BIBforeignlanguage{en}{Recent contributions to the mathematical
  theory of communication},'' p.~12.

\bibitem{EXKSC}
G.~Xin and P.~Fan, ``Exk-sc: A semantic communication model based on
  information framework expansion and knowledge collision,'' \emph{Entropy},
  vol.~24, no.~12, 2022.

\bibitem{zhou_adaptive_2022}
Q.~Zhou \emph{et~al.}, ``\BIBforeignlanguage{en}{Adaptive {Bit} {Rate}
  {Control} in {Semantic} {Communication} {With} {Incremental}
  {Knowledge}-{Based} {HARQ}},'' \emph{\BIBforeignlanguage{en}{IEEE Open J.
  Commun. Soc.}}, vol.~3, pp. 1076--1089, 2022.

\bibitem{lu_rethinking_2023}
K.~Lu \emph{et~al.}, ``Rethinking {Modern} {Communication} from {Semantic}
  {Coding} to {Semantic} {Communication},'' \emph{IEEE Wireless Commun.},
  vol.~30, no.~1, pp. 158--164, Feb. 2023.

\bibitem{wang_knowledge_2023}
B.~Wang \emph{et~al.}, ``Knowledge {Enhanced} {Semantic} {Communication}
  {Receiver},'' \emph{IEEE Commun. Lett.}, pp. 1--1, 2023.

\bibitem{farsad_deep_2018}
N.~Farsad, M.~Rao, and A.~Goldsmith, ``Deep learning for joint source-channel
  coding of text,'' in \emph{Proc. {IEEE} {ICASSP}}, Calgary, AB, Canada, 2018,
  pp. 2326--2330.

\bibitem{zhang_toward_2022}
P.~Zhang, W.~Xu, H.~Gao, K.~Niu, X.~Xu, X.~Qin, C.~Yuan, Z.~Qin, H.~Zhao,
  J.~Wei, and F.~Zhang, ``\BIBforeignlanguage{zh}{Toward
  {Wisdom}-{Evolutionary} and {Primitive}-{Concise} {6G}: {A} {New} {Paradigm}
  of {Semantic} {Communication} {Networks}},''
  \emph{\BIBforeignlanguage{zh}{Engineering}}, vol.~8, pp. 60--73, Jan. 2022.

\bibitem{yang_semantic_2022}
\BIBentryALTinterwordspacing
W.~Yang, H.~Du, Z.~Liew, W.~Y.~B. Lim, Z.~Xiong, D.~Niyato, X.~Chi, X.~S. Shen,
  and C.~Miao, ``\BIBforeignlanguage{en}{Semantic {Communications} for {6G}
  {Future} {Internet}: {Fundamentals}, {Applications}, and {Challenges}},''
  Jun. 2022, arXiv:2207.00427. [Online]. Available:
  \url{http://arxiv.org/abs/2207.00427}
\BIBentrySTDinterwordspacing

\bibitem{gunduz_beyond_2022}
D.~Gündüz \emph{et~al.}, ``Beyond transmitting bits: Context, semantics, and
  task-oriented communications,'' \emph{IEEE J. Sel. Areas Commun.}, vol.~41,
  no.~1, pp. 5--41, 2023.

\bibitem{wang_performance_2022}
Y.~Wang \emph{et~al.}, ``\BIBforeignlanguage{en}{Performance {Optimization} for
  {Semantic} {Communications}: {An} {Attention}-based {Reinforcement}
  {Learning} {Approach}},'' \emph{\BIBforeignlanguage{en}{IEEE J. Select. Areas
  Commun.}}, pp. 1--1, 2022.

\bibitem{dai_nonlinear_2022}
J.~Dai, S.~Wang, K.~Tan, Z.~Si, X.~Qin, K.~Niu, and P.~Zhang, ``Nonlinear
  {Transform} {Source}-{Channel} {Coding} for {Semantic} {Communications},''
  \emph{IEEE J. Sel. Areas Commun.}, vol.~40, no.~8, pp. 2300--2316, 2022.

\bibitem{shao_learning_2022}
J.~Shao, Y.~Mao, and J.~Zhang, ``\BIBforeignlanguage{en}{Learning
  {Task}-{Oriented} {Communication} for {Edge} {Inference}: {An} {Information}
  {Bottleneck} {Approach}},'' \emph{\BIBforeignlanguage{en}{IEEE J. Select.
  Areas Commun.}}, vol.~40, no.~1, pp. 197--211, Jan. 2022.

\bibitem{shao_task-oriented_2022}
J.~Shao, Y.~Mao, and J.~Zhang, ``Task-oriented communication for multidevice
  cooperative edge inference,'' \emph{IEEE Trans. Wireless Commun.}, vol.~22,
  no.~1, pp. 73--87, 2023.

\bibitem{xie_robust_2022}
\BIBentryALTinterwordspacing
S.~Xie, Y.~Wu, S.~Ma, M.~Ding, Y.~Shi, and M.~Tang,
  ``\BIBforeignlanguage{en}{Robust {Information} {Bottleneck} for
  {Task}-{Oriented} {Communication} with {Digital} {Modulation}},'' Sep. 2022,
  arXiv:2209.10382v1. [Online]. Available:
  \url{https://arxiv.org/abs/2209.10382v1}
\BIBentrySTDinterwordspacing

\bibitem{tang_contrastive_2023}
\BIBentryALTinterwordspacing
S.~Tang \emph{et~al.}, ``Contrastive {Learning} based {Semantic}
  {Communication} for {Wireless} {Image} {Transmission},'' Apr. 2023,
  arXiv:2304.09438. [Online]. Available: \url{http://arxiv.org/abs/2304.09438}
\BIBentrySTDinterwordspacing

\bibitem{xie_lite_2021}
H.~Xie and Z.~Qin, ``\BIBforeignlanguage{en}{A {Lite} {Distributed} {Semantic}
  {Communication} {System} for {Internet} of {Things}},''
  \emph{\BIBforeignlanguage{en}{IEEE J. Select. Areas Commun.}}, vol.~39,
  no.~1, pp. 142--153, Jan. 2021.

\bibitem{xie_deep_2021}
H.~Xie \emph{et~al.}, ``\BIBforeignlanguage{en}{Deep {Learning} {Enabled}
  {Semantic} {Communication} {Systems}},'' \emph{\BIBforeignlanguage{en}{IEEE
  Trans. Signal Process.}}, vol.~69, pp. 2663--2675, 2021.

\bibitem{zhou_semantic_2022}
Q.~Zhou, R.~Li, Z.~Zhao, C.~Peng, and H.~Zhang,
  ``\BIBforeignlanguage{en}{Semantic {Communication} {With} {Adaptive}
  {Universal} {Transformer}},'' \emph{\BIBforeignlanguage{en}{IEEE Wireless
  Commun. Lett.}}, vol.~11, no.~3, pp. 453--457, Mar. 2022.

\bibitem{xie_task-oriented_2022}
H.~Xie, Z.~Qin, X.~Tao, and K.~B. Letaief, ``Task-{Oriented} {Multi}-{User}
  {Semantic} {Communications},'' \emph{IEEE J. Select. Areas Commun.}, vol.~40,
  no.~9, pp. 2584--2597, Sep. 2022.

\bibitem{yang_ofdm-guided_2022}
M.~Yang, C.~Bian, and H.-S. Kim, ``\BIBforeignlanguage{en}{{OFDM}-{Guided}
  {Deep} {Joint} {Source} {Channel} {Coding} for {Wireless} {Multipath}
  {Fading} {Channels}},'' \emph{\BIBforeignlanguage{en}{IEEE Trans. Cogn.
  Commun. Netw.}}, vol.~8, no.~2, pp. 584--599, Jun. 2022.

\bibitem{han_semantic-preserved_2023}
T.~Han, Q.~Yang, Z.~Shi, S.~He, and Z.~Zhang, ``Semantic-{Preserved}
  {Communication} {System} for {Highly} {Efficient} {Speech} {Transmission},''
  \emph{IEEE J. Select. Areas Commun.}, vol.~41, no.~1, pp. 245--259, Jan.
  2023.

\bibitem{huang_toward_2023}
D.~Huang, F.~Gao, X.~Tao, Q.~Du, and J.~Lu, ``Toward {Semantic}
  {Communications}: {Deep} {Learning}-{Based} {Image} {Semantic} {Coding},''
  \emph{IEEE J. Select. Areas Commun.}, vol.~41, no.~1, pp. 55--71, Jan. 2023.

\bibitem{jiang_wireless_2023}
P.~Jiang, C.-K. Wen, S.~Jin, and G.~Y. Li, ``Wireless {Semantic}
  {Communications} for {Video} {Conferencing},'' \emph{IEEE J. Select. Areas
  Commun.}, vol.~41, no.~1, pp. 230--244, Jan. 2023.

\bibitem{yue_learned_2023}
W.~Yue \emph{et~al.}, ``Learned {Source} and {Channel} {Coding} for
  {Talking}-{Head} {Semantic} {Transmission},'' in \emph{Proc. {IEEE} {WCNC}},
  Mar. 2023, pp. 1--6.

\bibitem{zhang_semantic_2023}
\BIBentryALTinterwordspacing
B.~Zhang, Z.~Qin, and G.~Y. Li, ``Semantic {Communications} with
  {Variable}-{Length} {Coding} for {Extended} {Reality},'' Mar. 2023,
  arXiv:2302.08645. [Online]. Available: \url{http://arxiv.org/abs/2302.08645}
\BIBentrySTDinterwordspacing

\bibitem{chen_trust-worthy_2023}
\BIBentryALTinterwordspacing
J.~Chen \emph{et~al.}, ``Trust-{Worthy} {Semantic} {Communications} for the
  {Metaverse} {Relying} on {Federated} {Learning},'' May 2023,
  arXiv:2305.09255. [Online]. Available: \url{http://arxiv.org/abs/2305.09255}
\BIBentrySTDinterwordspacing

\bibitem{zhang_deep_2022}
\BIBentryALTinterwordspacing
H.~Zhang \emph{et~al.}, ``\BIBforeignlanguage{en}{Deep {Learning}-{Enabled}
  {Semantic} {Communication} {Systems} with {Task}-{Unaware} {Transmitter} and
  {Dynamic} {Data}},'' Aug. 2022, arXiv:2205.00271. [Online]. Available:
  \url{http://arxiv.org/abs/2205.00271}
\BIBentrySTDinterwordspacing

\bibitem{devlin_bert_2019}
\BIBentryALTinterwordspacing
J.~Devlin \emph{et~al.}, ``{BERT}: {Pre}-training of {Deep} {Bidirectional}
  {Transformers} for {Language} {Understanding},'' May 2019, arXiv:1810.04805.
  [Online]. Available: \url{http://arxiv.org/abs/1810.04805}
\BIBentrySTDinterwordspacing

\bibitem{Simonyan:2014cmh}
\BIBentryALTinterwordspacing
K.~Simonyan and A.~Zisserman, ``Very {Deep} {Convolutional} {Networks} for
  {Large}-{Scale} {Image} {Recognition},'' Apr. 2015. [Online]. Available:
  \url{http://arxiv.org/abs/1409.1556}
\BIBentrySTDinterwordspacing

\bibitem{transformer_2017}
A.~Vaswani, N.~Shazeer, N.~Parmar, J.~Uszkoreit, L.~Jones, A.~N. Gomez,
  L.~Kaiser, and I.~Polosukhin, ``Attention is all you need,'' ser.
  NIPS'17.\hskip 1em plus 0.5em minus 0.4em\relax Red Hook, NY, USA: Curran
  Associates Inc., 2017, p. 6000–6010.

\bibitem{goodfellow_generative_2014}
I.~Goodfellow, J.~Pouget-Abadie, M.~Mirza, B.~Xu, D.~Warde-Farley, S.~Ozair,
  A.~Courville, and Y.~Bengio, ``Generative {Adversarial} {Nets},'' in
  \emph{Proc. {NeurIPS}}, 2014.

\bibitem{shannon_mathematical_nodate}
C.~Shannon and W.~Weaver, ``\BIBforeignlanguage{en}{The {Mathematical} {Theory}
  of {Communication}},'' \emph{\BIBforeignlanguage{en}{The University of
  Illinois Press}}, p. 131, 1949.

\bibitem{ramesh_hierarchical_2022}
\BIBentryALTinterwordspacing
A.~Ramesh, P.~Dhariwal, A.~Nichol, C.~Chu, and M.~Chen,
  ``\BIBforeignlanguage{en}{Hierarchical {Text}-{Conditional} {Image}
  {Generation} with {CLIP} {Latents}},'' Apr. 2022. [Online]. Available:
  \url{https://arxiv.org/abs/2204.06125v1}
\BIBentrySTDinterwordspacing

\bibitem{liu_rate-distortion_2021}
J.~Liu \emph{et~al.}, ``A {Rate}-{Distortion} {Framework} for {Characterizing}
  {Semantic} {Information},'' in \emph{Proc. {IEEE} {ISIT}}, Jul. 2021, pp.
  2894--2899.

\bibitem{sagduyu_is_2022}
\BIBentryALTinterwordspacing
Y.~E. Sagduyu, T.~Erpek, S.~Ulukus, and A.~Yener, ``\BIBforeignlanguage{en}{Is
  {Semantic} {Communications} {Secure}? {A} {Tale} of {Multi}-{Domain}
  {Adversarial} {Attacks}},'' Dec. 2022. [Online]. Available:
  \url{https://arxiv.org/abs/2212.10438v1}
\BIBentrySTDinterwordspacing

\bibitem{huang_deep_2021}
D.~Huang, X.~Tao, F.~Gao, and J.~Lu, ``Deep {Learning}-{Based} {Image}
  {Semantic} {Coding} for {Semantic} {Communications},'' in \emph{Proc. IEEE
  {GLOBECOM}}.\hskip 1em plus 0.5em minus 0.4em\relax Madrid, Spain: IEEE, Dec.
  2021, pp. 1--6.

\bibitem{lu_reinforcement_2022}
\BIBentryALTinterwordspacing
K.~Lu, R.~Li, X.~Chen, Z.~Zhao, and H.~Zhang,
  ``\BIBforeignlanguage{en}{Reinforcement {Learning}-powered {Semantic}
  {Communication} via {Semantic} {Similarity}},'' Apr. 2022, arXiv:2108.12121.
  [Online]. Available: \url{http://arxiv.org/abs/2108.12121}
\BIBentrySTDinterwordspacing

\bibitem{Europarl_2005}
P.~Koehn, ``\BIBforeignlanguage{en}{Europarl: A parallel corpus for statistical
  machine translation},'' \emph{\BIBforeignlanguage{en}{MT summit}}, vol.~5,
  pp. 79--86, Sep. 2005.

\bibitem{ronneberger_u-net_2015}
\BIBentryALTinterwordspacing
O.~Ronneberger, P.~Fischer, and T.~Brox, ``U-{Net}: {Convolutional} {Networks}
  for {Biomedical} {Image} {Segmentation},'' May 2015, arXiv:1505.04597.
  [Online]. Available: \url{http://arxiv.org/abs/1505.04597}
\BIBentrySTDinterwordspacing

\bibitem{glorot_deep_2010}
X.~Glorot, A.~Bordes, and Y.~Bengio, \emph{Deep {Sparse} {Rectifier} {Neural}
  {Networks}}, Jan. 2010, vol.~15, journal of Machine Learning Research.

\bibitem{maas_rectier_nodate}
A.~L. Maas, A.~Y. Hannun, and A.~Y. Ng, ``\BIBforeignlanguage{en}{Rectiﬁer
  {Nonlinearities} {Improve} {Neural} {Network} {Acoustic} {Models}}.''

\bibitem{Kingma:2014vow}
\BIBentryALTinterwordspacing
D.~P. Kingma and J.~Ba, ``Adam: {A} {Method} for {Stochastic} {Optimization},''
  Jan. 2017, arXiv:1412.6980. [Online]. Available:
  \url{http://arxiv.org/abs/1412.6980}
\BIBentrySTDinterwordspacing

\bibitem{elias1955coding}
P.~Elias, ``Coding for noisy channels,'' in \emph{RE Conv. Rec.}, 1955, pp.
  37--46.

\bibitem{papineni_bleu_2002}
K.~Papineni \emph{et~al.}, ``Bleu: a {Method} for {Automatic} {Evaluation} of
  {Machine} {Translation},'' in \emph{Proc. {ACL'02}}, Jul. 2002, pp. 311--318.

\bibitem{laurens_visualizing_2008}
V.~D.~M. Laurens \emph{et~al.}, ``Visualizing {Data} using t-{SNE},''
  \emph{Journal of Machine Learning Research}, vol.~9, no. 2605, pp.
  2579--2605, 2008.

\end{thebibliography}
\bibliographystyle{IEEEtran}

\end{document}